\documentclass[structabstract]{aa}  
\usepackage{graphicx}
\usepackage{txfonts}
\usepackage{natbib}
\begin{document}
\title{Single peaked CO emission line profiles from the inner regions of protoplanetary 
disks 
\thanks{This work is based on observations collected at the European Southern 
Observatory Very Large Telescope under program ID 179.C-0151.}}

\author{J.E. Bast\inst{1}
  \and    
  J.M. Brown\inst{2}
  \and{G.J. Herczeg}\inst{2}
   \and
  E.F. van Dishoeck\inst{1,2}
  \and
  K.M. Pontoppidan\inst{3}   
}

\institute{Leiden Observatory, Leiden University,
  P.O. Box 9513, 2300 RA Leiden, The Netherlands
  \email{bast@strw.leidenuniv.nl}
  \and
  Max Planck Institute for Extraterrestrial Physics, Giessenbachstrasse 1, 85748 Garching, Germany
  \and
  Division of Geology and Planetary Science, Mail code 150-21, California Institute of Technology, 
  Pasadena, CA 91125, USA  
}

\date{Received month date, year; accepted month date, year}

\abstract 
{Protoplanetary disks generally exhibit strong line emission
  from the CO fundamental v=1--0 ro-vibrational band around
  4.7\,$\mu$m. The lines are usually interpreted as being formed in
  the Keplerian disk, as opposed to other kinematic components of the
  young stellar system.}  {This paper investigates a set of disks that
  show CO emission line profiles characterized by a single, narrow
  peak and a broad base extending to $>50\,\rm km\,s^{-1}$,
  not readily explained by just Keplerian motions of gas in
    the inner disk.}  {High resolution ($R$ = 10$^5$) $M$-band
  spectroscopy has been obtained using CRIRES at the Very Large
  Telescope in order to fully resolve fundamental ro-vibrational CO
  emission line profiles around 4.7 $\mu$m.}  
{Line profiles with a narrow peak and broad wings are found for 8 disks among a sample of $\sim$50 disks around T Tauri stars with CO
  emission. The lines are very symmetric, have high line/continuum
  ratios and have central velocity shifts of $<$ 5 km\,s$^{-1}$
  relative to the stellar radial velocity. The disks in this
  subsample are accreting onto their central stars at high rates
  relative to the parent sample. All 8 disks show CO emission
    lines from the $v=2$ vibrational state and 4/8 disks show emission
    up to $v = 4$. Excitation analyses of the integrated line fluxes
  reveal a significant difference between typical rotational
  ($\sim$300-800 K) and vibrational ($\sim$1700 K) temperatures,
  suggesting that the lines are excited, at least in part, by
  UV-fluorescence. For at least one source, the narrow and
    broad components show different excitation temperatures, but
    generally the two component fits have similar central velocities
    and temperature.  Analysis of their spatial distribution shows
    that the lines are formed within a few AU of the central star.}
      {It is concluded that these broad centrally peaked line
    profiles are inconsistent with the double peaked profiles expected
    from just an inclined disk in Keplerian rotation. Models in which the low velocity emission arises from large disk radii are excluded based
    on the small spatial distribution. Alternative non-Keplerian line
    formation mechanisms are discussed, including thermally and
    magnetically launched winds and funnel flows. The most likely
    interpretation is that the broad-based centrally peaked line
    profiles originate from a combination of emission from the inner
    part ($<$ a few AU) of a circumstellar disk, perhaps with enhanced
    turbulence, and a slow moving disk wind, launched by either EUV
    emission or soft X-rays.}

  
  \authorrunning{J.Bast et al.}
  \titlerunning{Single peaked CO emission line profiles from inner regions of disks}
  
  \maketitle
  %
%________________________________________________________________

\section{Introduction}

It is generally thought that planets form in the inner regions of
protoplanetary disks \citep[$\lesssim$10
AU][]{Lissauer1993}. Information on the physical structure, gas
dynamics and chemical composition of the planet-forming region is
essential to constrain models of planet formation. Processes like
planet migration, which can change the orbits of newly formed planets,
depend sensitively on the presence of gas in the disk
\citep[e.g.,][]{Ward1997, Kley09}. The planetary mass distributions
\citep{Ida04} and planetary orbits \citep{Kominami02, Trilling02}
resulting from planet formation models can eventually be tested
against recent observations of exo-planetary systems
\citep[e.g.,][]{Mordasini091, Mordasini092}. Observations of line emission from
disks at high spectral and spatial resolution are needed to provide
the initial conditions for these models. Gas-phase tracers of the
disk surface can also be used to probe photo-evaporation processes
 \citep{Gorti09_1,Gorti09_2}. More generally, these observations provide
constraints on the lifetime of the gas in the inner part of the disk
and thus its ability to form giant gaseous planets.

The bulk of the gas mass in protoplanetary disks is in H$_2$ but this
molecule is difficult to observe since its rotational quadrupole
transitions from low-energy levels are intrinsically weak and lie in
wavelength ranges with no or poor atmospheric transmission
\citep[e.g.,][]{Carmona07}. In contrast, the next most
abundant molecule, CO, has ro-vibrational lines which can be readily
detected from the ground. This makes CO an optimal tracer of the
characteristics of the warm gas in the inner regions of 
disks \citep[see][for overview]{Najita07}. CO overtone emission ($\Delta{\rm{v}}$ = 2) was detected for the
first time in low and high mass young stellar objects by
\citet{Thompson1985} and was attributed to circumstellar disks by
\citet{Carr1989}.

Overtone emission lines at 2.3 $\mu$m from disks around T Tauri stars, when
present, have been fitted with double peaked line profiles with a
FWHM of around 100 km s$^{-1}$, which suggests an origin in the
innermost part (0.05-0.3 AU) of the disk under the assumption of
Keplerian rotation \citep{Carr1993, Chandler1993, Najita1996}. The
relative intensities of the ro-vibrational lines can be used to
determine characteristic CO excitation temperatures (rotational and
vibrational), which, in turn, provide constraints on the kinetic temperatures and
densities in the line-forming region. The overtone data resulted in temperature estimates 
in the 1500-4000 K range and with densities of $>$10$^{10}$ cm$^{-3}$
\citep{Chandler1993, Najita00, Najita07}.

Fundamental CO $\rm{v} = 1-0$ emission at 4.7 $\mu$m has been observed
both from disks around T Tauri stars \citep[e.g.,][]{Najita03,
Rettig04, Salyk07, Pontoppidan08, Salyk09} and around Herbig Ae/Be
stars \citep[e.g.,][]{Brittain03, Blake04, Brittain07, Brittain09,
Plas09}. The fundamental ro-vibrational lines are excited at lower
temperatures (1000-1500 K) than the overtone lines \citep{Najita07}.
Fundamental CO emission lines are usually fitted with double peaked or
narrow single-peaked line profiles that can be described by a
Keplerian model. Single-peaked line profiles with broad wings (up to
100 km s$^{-1}$) in T Tauri disks have been seen by
\citet{Najita03}, who also discussed their origin. Three possible formation
scenarios were mentioned: disk winds, funnel flows or gas in the
rotating disk. The latter option was favored, and the lack of a double peak was ascribed to the relatively low ($R$=$\lambda$/$\Delta\lambda$=25000) spectral resolving power of the data. Alternative explanations such as an origin in disk winds and
funnel flows were ruled out mainly because of the lack of asymmetry in
the line profiles.

The high resolution ($R$ = 10$^5$) spectrometer CRIRES (CRyogenic
InfraRed Echelle Spectrograph) fed by the MACAO (Multi - Application
Curvature Adaptive Optics) adaptive optics system on the Very Large
Telescope offers the opportunity to observe molecular gas emission
from T-Tauri disks with unsurpassed spectral and spatial resolution.
A sample of $\sim$70 disks was observed in the fundamental CO band
around 4.7\,$\mu$m as part of an extensive survey of molecular
emission from young stellar objects. In total, 12 of the 70 T Tauri
stars show CO emission lines with a broad base and a narrow central
peak, from now on called broad-based single peaked line
  profiles. Eight of the 12 T Tauri stars are selected for detailed
analysis in this paper, based on criteria discussed in \S
\ref{source_selection}. Because the lines remain single peaked even
when observed at 4 times higher spectral resolution than previous
observations, the lack of a double peak can no longer be explained by
the limited resolution in \citet{Najita03}. Hence the modeled double
peaked lines in \citet{Najita03} are not a plausible explanation for
the centrally peaked line profiles in T Tauri disks. The aim of this
paper is to classify these broad centrally peaked lines and to
constrain their origin. Since many of these sources have high line to
continuum ratios and are prime targets to search for molecules other
than CO \citep[e.g.][]{Salyk08}, a better understanding of these
sources is also warranted from the perspective of disk chemistry
studies.

The observations and sample are presented in $\S$\ref{obs}. In
  section $\S$\ref{profile} the line profiles are modeled using a
  Keplerian disk model where it is concluded that a model with a
  standard power-law temperature structure does not provide a good fit
  to the broad-based single peaked line profiles. The profiles are
  subsequently inverted to determine what temperature distribution
  would be consistent with the spectra. The origin of the emission is
  then further constrained in $\S$\ref{char} by extracting radial
  velocity shifts between the gas and the star, determining rotational
  and vibrational temperatures and investigating the extent of the
  emission. In $\S$\ref{discussion} the results are discussed and
they are summarized in \S\ref{conclusions}.

  \begin{table*}
\footnotesize
\caption{List of sources with single-peaked line profiles analyzed in this paper.}
 \begin{minipage}[t]{\columnwidth}
 \renewcommand{\footnoterule}{}  
 \centering
  \thispagestyle{empty}
  \begin{tabular}{l l r c c c c c l}
  \hline
   \hline
   Source\footnote{Three of these sources are binaries and the separations between their A and B components are: AS 205: 1\farcs3, S CrA: 1\farcs3 and VV CrA: 1\farcs9 \citep{Reipurth93}.} & $\alpha$(J2000) & $\delta$(J2000) & Spectral Type & Distance (pc) & Flux 
   [Jy] & $T_{\rm{eff}}$ [K] & $L_\star$ [$L_\odot$] & Ref.\footnote{References. - (1) \citet{Kessler06}; (2) 
   \citet{Prato03}; (3) \citet{Takami03}; (4) \citet{Salyk08}; (5) \citet{Muzerolle03}; (6) \citet{Guenther08}, (7) \citet{Luhman08}, (8) \citet{Evans09}, 
   (9) \citet{Whittet1997}; (10) \citet{Andrews09}; (11) \citet{Ricci10}; (12) \citet{Gras05}; (13) \citet{Koresko1997}; (14) \citet{Natta00}; (15) \citet{Stempels03} (16) \citet{Appenzeller86} and (17) Peterson et al. (subm.).} \\
   \hline  
AS 205 A (N)\footnote{N or S indicate if the source is the northern (N) or southern (S) of a binary system.} & 16 11 31.4 & -18 38 24.5 & K5 & 125 & 4.0 & 4250 & 4.0 & 1, 4, 8, 10 \\          
DR Tau & 04 47 06.2 & +16 58 42.9 & K7 & 140 & 1.3 &  4060 & 1.1 & 4, 5, 11 \\     
RU Lup  & 15 56 42.3 & -37 49 15.5 & K7 & 140 & 1.1 & 4000 & 1.3 & 1, 6, 12, 15 \\                                                         
S CrA A (N) & 19 01 08.6 & -36 57 20.0 & K3 & 130 & 2.2 & 4800 & 2.3,  & 2, 3, 17 \\  
S CrA B (S) & 19 01 08.6 & -36 57 20.0 & M0 & 130 & 0.8 & 3800 & 0.8 & 2, 3, 17 \\    
VV CrA A (S) & 19 03 06.7 & -37 12 49.7 & K7 & 130 & - & 4000 & 0.3 & 3, 13, 16 \\
VW Cha & 11 08 01.8 & -77 42 28.8 & K5 & 178 & 0.7 & 4350 & 2.9 & 7, 9, 14 \\  
VZ Cha & 11 09 23.8 & -76 23 20.8 & K6 & 178 & 0.4 & 4200 & 0.5 & 7, 9, 14 \\
  \hline
\end{tabular} 
 \end{minipage}
\label{tab:parameters1}
\end{table*}

   \begin{table*}
\footnotesize
\caption{Journal of observations.}
 \begin{minipage}[t]{\columnwidth}
  \renewcommand{\footnoterule}{} 
 \centering
  \thispagestyle{empty}
  \begin{tabular}{l l l l l l}
   \hline
   \hline
   Source & Obs. time & Settings ($\mu$m)\footnote{The reference 
   wavelength of a given setting is centered on the third detector.} &
    Standard star & Spectral type \\
      &&&& standard star \\
   \hline                             
AS 205 A & Apr 07 & 4.760, 4.662, 4.676, 4.773  & BS 4757 & A0 \\ 
 & Aug 07 & 4.730 & BS 5812 & B2.5 \\ 
 & Apr 08 & 4.730 & BS 6084 & B1 \\  
 & Aug 09 & 5100, 5115 & BS 5984 & B0.5 \\                               
DR Tau & Oct  07 & 4.716, 4.730, 4.833, 4.868 & BS 3117, BS 838 & B3, B8 \\   
 & Dec 08 & 4.716, 4.946 & BS 1791 & B7 \\                 	 
RU Lup  & Apr 07 & 4.716, 4.730, 4.833, 4.929 & BS 5883 & B9 \\  	
 & Apr 08 & 4.730 & BS 6084 & B1 \\       
S CrA & Apr 07 & 4.730, 4.716 & BS 6084 & B1 \\ 
 & Aug 07 & 4.730 & BS 7235, BS 7920 & A0, A9 \\   
 & Aug 08 & 4.868, 4.946 & BS 6084, BS 7236 & B1, B9 \\ 	                  
VV CrA A & Apr 07 & 4.716, 4.730, 4.840 & BS 7362 & A4 \\
 & Aug 07 & 4.770, 4.779 & BS 7236 & B9 \\
 & Aug 08 & 4.946 & BS 7236 & B9 \\
VW Cha & Dec 08 & 4.716, 4.800, 4.820, 4.946 & BS 5571, HR 4467 & B2, B9 \\
VZ Cha & Dec 08 & 4.716, 4.800, 4.820, 4.946 & BS 5571, HR 4467 & B2, B9 \\
  \hline
\end{tabular}
 \end{minipage}
\label{tab:setting}
\end{table*}

\section{Observations and sample} \label{obs}
A sample of 70 disks around low-mass pre-main sequence stars was
 observed at high spectral resolving power ($\lambda/\Delta\lambda$ =
 10$^5$ or 3 km s$^{-1}$) with CRIRES mounted on UT1 at the Very Large
 Telescope (VLT) of the European Southern Observatory, Paranal,
 Chile. The CRIRES instrument \citep{Kaufl04} is fed by an adaptive
 optics system \citep[MACAO,][]{Paufique04}, resulting in a typical
 spatial resolution of $\sim$160-200 milli-arcsec along the
 slit. CRIRES has 4 detectors that each cover about 0.02 -- 0.03
 $\mu$m with gaps of about 0.006 $\mu$m at 4.7 $\mu$m. Staggered pairs of
 settings shifted in wavelength are observed to cover the detector gaps and
 produce continuous spectra.

 The observations were taken during a period from April 21 2007 to
 January 3 2009. The parent sample is a broad selection of low-mass
 young stellar objects, consisting mostly of T Tauri stars and a few
 Herbig Ae stars. The full data set will be published in a future
 study (Brown et al.\ in prep.). Of the 70 sources, about 50 disks
 around T Tauri stars show clear CO emission lines. This paper
 focuses on a subsample of 8 of the 12 objects that show broad-based
 single peaked CO ro-vibrational line profiles. Their names and
 characteristic parameters are presented in
 Table~\ref{tab:parameters1}, and their selection is justified in \S
 3.2 and \S 4. Inclinations are unknown for the majority of the
 sources. Several of the sources in the sample are binaries, with the
 primary and secondary defined as A and B, respectively. The specific
 definition for each source's primary and secondary is taken from the
 literature, see Table~\ref{tab:parameters1}. For S CrA,
   both A and B components show broad-based centrally peaked line
   profiles. For AS 205 and VV CrA, only the A component has such
   single-peaked emission profiles. The B components show CO
   absorption (see \citealt{Smith09} for the case of VV CrA B). Thus,
   there is no obvious trend of the presence of these profiles with
   binarity of the system.

Ground-based M-band spectroscopy is usually dominated by strong sky
emission lines (including CO itself) superimposed on a thermal
continuum. Nodding with a throw of 10" was performed to correct for
both the sky emission and the thermal continuum. In addition
jittering with a random offset within a radius of 0\farcs5 was used to
decrease systematics and to remove bad pixels. The weather conditions
during the observations were usually good with optical seeing varying
typically between 0\farcs5 -- 1\farcs5.  Even at times when the seeing
was higher most of the light still passed through the 0\farcs2 slit
because of the adaptive optics system. Observations of standard stars
were done close in time to each science target with airmass
differences of typically 0.05--0.1 in order to correct for telluric
features.

The observations typically covered CO $\rm{v} = 1-0$: R(8)--P(32)
lines, where the notation indicates that most lines between R(8) and
P(32) are observed, including, for example, the R(0) and P(1)
lines. Lines from vibrationally excited levels as well as
isotopologues are also included, specifically: CO $\rm{v} = 2-1$:
R(17)--P(26); CO $\rm{v} = 3-2$: R(26)--P (21); CO $\rm{v} = 4-3$:
R(37)--P(15); $^{13}$CO: R(24)--P(23); C$^{18}$O: R (17)--P(27); and
C$^{17}$O: R(26)--P(22).

The wavelengths and dates of the observations can be found in
Table~\ref{tab:setting}. Each wavelength setting is centered on the
third, in order of increasing wavelength, of the four detectors. As an
example, the setting at 4.730\,$\mu$m corresponds to a spectral range
of 4.660 -- 4.769 $\mu$m. Observations were sometimes taken during
different seasons to shift the telluric lines relative to the source
spectrum due to the reflex motion of the Earth.  Combination of the
two data sets then allows the reconstruction of the complete line
profile, whereas a single epoch would have a gap due to the presence of
a saturated telluric CO line. For 6 of the sources presented here,
no evidence has been found that the line fluxes and profiles changed
within this 1--1.5 yr period. For VW Cha and VZ Cha this information
is not available since they were only observed once.

\subsection{Data reduction}\label{reduction}

 \begin{figure*}
\centering
{
 \includegraphics[width=160mm, angle=0.0]{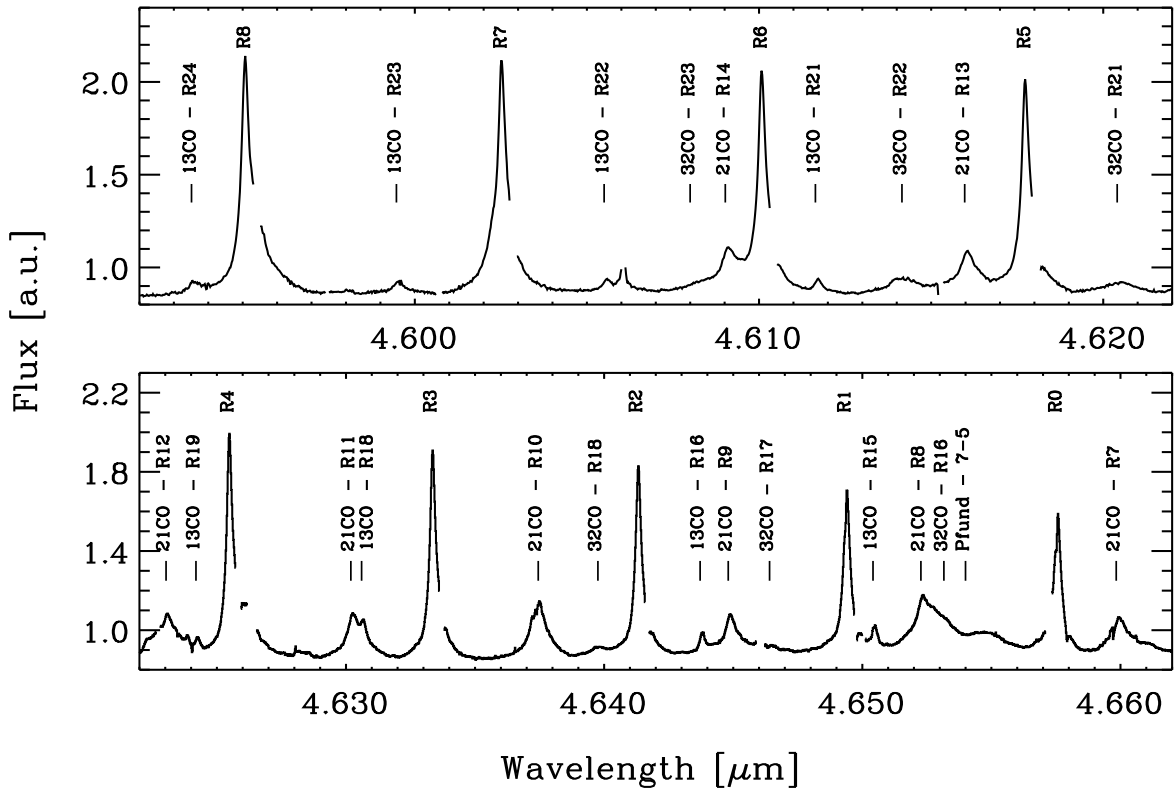}
 \caption{Part of the VLT-CRIRES spectrum of AS 205 A. The strong
$^{12}$CO lines are labelled with their $R$-band lower $J$-transition
number and the other marked lines show detections of $^{12}$CO $\rm{v}$ =
2--1, $\rm{v}$ = 3-2, $^{13}$CO, and the Pfund-$\beta$ line. The gaps
in the spectrum are excised telluric features. For other parts of the 
AS 205 A spectrum, these gaps have been filled in by observations at
different times of the year (see Fig.~\ref{fig:source_setting}).}
 \label{fig:onesetting}
}
\end{figure*} 

The spectra were reduced using standard methods for infrared
spectroscopy \citep[see][for details]{Pontoppidan08}. Dome flats were
used for each setting to correct the images for pixel-to-pixel
sensitivity differences in the detectors. A linearity correction was
applied to the frames using the parametrization given in the CRIRES
documentation. The nod pairs were then differenced to subtract the
background. The data stack was co-added after correction
for field distortion. Extraction of the 2-dimensional images to 1
dimensional spectra was done after the subtraction and combination of
the 2-D frames tracing the source in the 2-D spectra. Additional
spectra of well separated binaries were extracted as well during this
procedure.

The spectra were subsequently wavelength calibrated by fitting the
standard star atmospheric lines to an atmospheric model generated
using the Reference Forward Model (RFM) code. This model is a
  line-by-line radiative transfer model developed at Oxford University
  based on \citet{Clough82}\footnote{{\tt
      http://www.atm.ox.ac.uk/RFM/}}. The typical velocity accuracy
is about 0.1--1 km s$^{-1}$, but can vary between settings, depending
on the density of telluric lines. The last step in the reduction
process was to correct for the strong telluric absorption lines in the
spectra, by dividing the science spectra with the spectra taken of
early-type photospheric standard stars. These hot stars have a
  strong hydrogen Pfund $\beta$ (7-5) absorption line at 4.654
  $\mu$m. Division by the standard star therefore introduces an
  artificial contribution to the emission line in the science
  spectra. However, Pfund $\beta$ is detected in emission towards most
  of the sources prior to telluric correction. Early-type
photospheric standard stars are otherwise rather featureless
throughout the spectral band. Flux calibration was carried out by
scaling to photometry from the Spitzer IRAC band 2 (see Table
\ref{tab:parameters1}).

 \begin{figure*}
\centering
{
 \includegraphics[width=160mm]{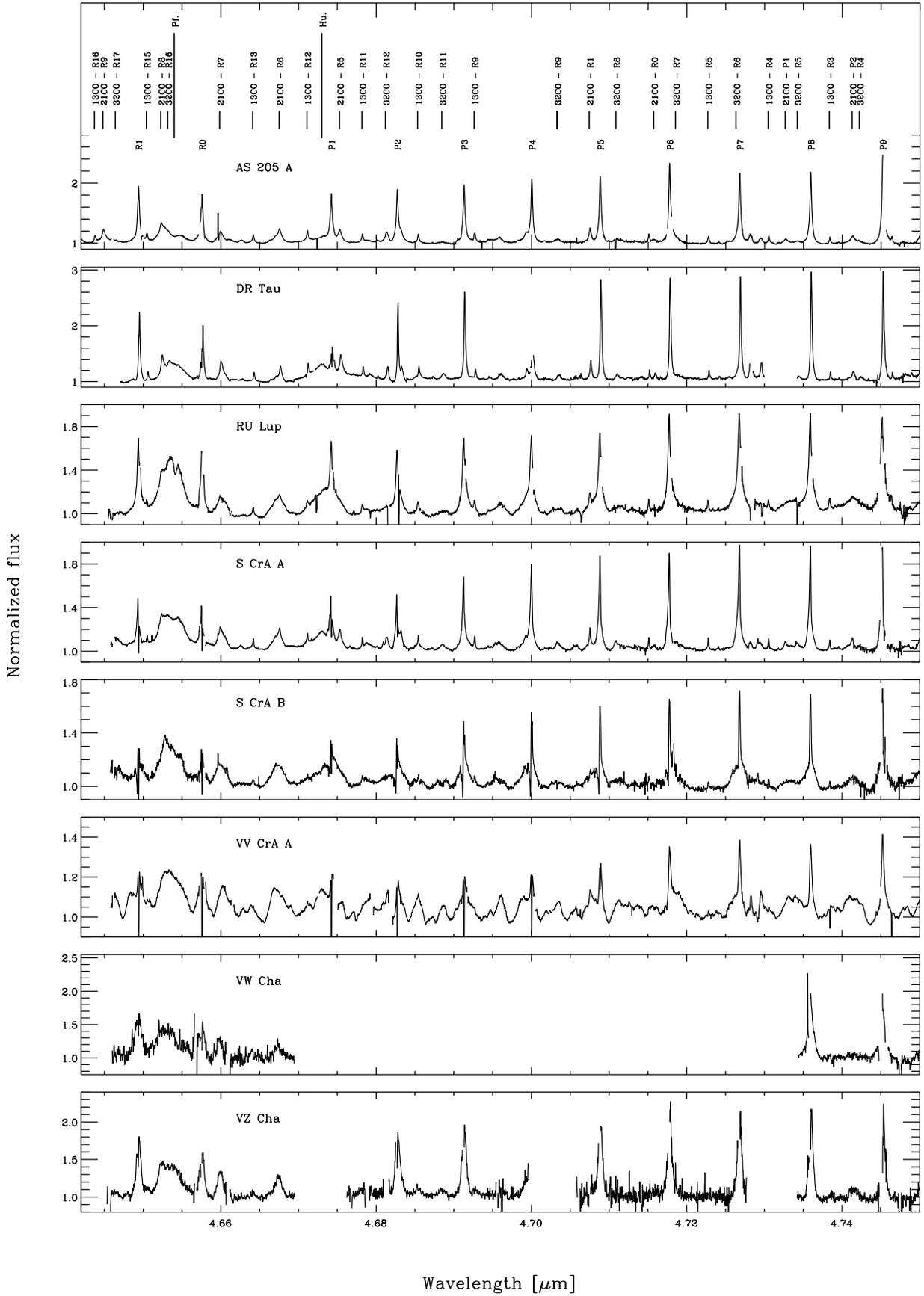}
 \caption{Partial spectra for the entire sample of sources with single
peaked $^{12}$CO lines. The HI Pfund $\beta$ line at 4.655 $\mu$m is
included. The spikes seen in spectra of S CrA A, B and VV CrA A are
narrow (self) absorption lines visible in the low $J$ lines.}
\label{fig:source_setting}
}
\end{figure*}

In Fig.~\ref{fig:onesetting} part of the reduced spectrum for the northern of the binary components of AS 205 is presented. This spectrum shows
highly resolved $^{12}$CO line profiles for the ${\rm v}=1-0$ transitions
R(8)--R(0) and clear detections of $^{13}$CO v$=1-0$, CO $\rm{v} =
2-1$ and $\rm{v} = 3-2$ lines and the Pfund $\beta$ line. An
overview of parts of the spectra of the 8 single peaked sources is
presented in Fig.~\ref{fig:source_setting}.

\section{Line profiles} \label{profile}

\subsection{$^{12}$CO line profiles}

\begin{figure*}[htb]
\centering
{
 \includegraphics[width=160mm]{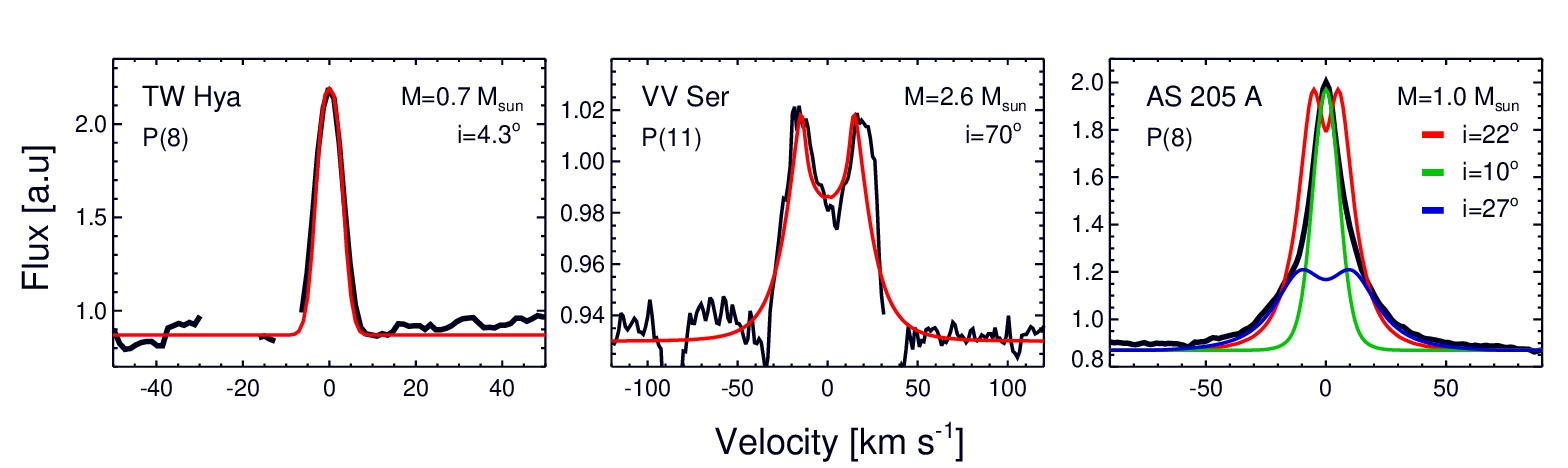}
 \caption{Modeled lines, plotted in red, are fitted to observed
 spectral lines in black of TW Hya, VV Ser and AS 205 A. Good fits
 could be found for VV Ser and TW Hya using a Keplerian model with a power-law temperature profile. No good fits could be found for AS 205
 A. The optimal fit to the total line profile (red) using extended
 emission out to 10 AU, to the narrow inner part (green) and to the
 outer broader wings (blue) are presented in the right figure. See
 text for model parameters.}
 \label{fig:linemodelsum}
}
\end{figure*}

The spectrally resolved fundamental CO emission lines from the
sample of T Tauri stars observed with CRIRES show a broad variety of
profiles. Often, the line profiles can be explained by Keplerian
rotation, absorption by the disk or absorption by a foreground cloud (Brown et al., in prep). Fig.~\ref{fig:linemodelsum} shows a selection of
emission line profiles of three basic shapes: A) narrow single-peaked,
matched by a single Gaussian, B) double-peaked line profiles and C)
single-peaked with broad wings. Lines that show absorption are not
included due to the difficulty in determining the profile close to line
center.

The CO line profiles of TW Hya fall in category A and likely arise
from gas in a face-on disk in Keplerian rotation \citep[e.g.][]
{Pontoppidan08}. Prototypical for category C are the line profiles of the source AS 205 A, which are by inspection qualitatively different
from the double peaked profiles of sources like AA Tau and VV Ser (category B) because of their single narrow peak relative to a broad base. While categories A and B appear to be well understood (see below), the question is what
the origin of the broad centrally peaked line profiles of category C
is.

\subsection{Keplerian disk model}\label{keplerian}

To investigate to what extent the broad
centrally peaked (category C) line profiles can be described by
emission originating from gas circulating in the inner parts of the
disk, a simple disk model of gas in Keplerian rotation adopted from
\cite {Pontoppidan08} is used. 

\subsubsection{Description of the model}\label{model}
 
The model includes a flat disk with constant radial surface
density. The motions of the gas in the disk are described by Kepler's
law. The disk is divided into emitting rings with radial sizes that
are increasing logarithmically with radius. Each ring has a specific
temperature. Local thermal equilibrium (LTE) is used as an
approximation for describing the excitation of the emitting molecules,
which means that the relative populations in a ro-vibrational level
can be derived using the Boltzmann distribution at the given
temperature in each emitting ring.

In the standard formulation, a power law is used to describe
the temperature gradient throughout the disk,
$T=T_{0}(R/R_{0})^{-\alpha}$, where $T$ is the gas temperature, $R$ is
the radius and $T_{0}$ is the temperature at the inner radius
$R_{0}$. The other parameters are: the mass of the star $M_\star$; the
inclination of the disk $i$; the outer radius of the disk
$R_{\rm{max}}$; and a line broadening parameter $\Delta{V}_{\rm{mod}}$
which combines the instrumental broadening and any turbulent
broadening. Several parameters are degenerate, including the stellar
mass and disk inclination. Thus, this model is not used to derive the
best estimates of the characteristic parameters of the sources. The
aim is instead to explore the parameter space to see whether good fits
can be found for the category C line profiles using basic Keplerian
physics. The local line profile is convolved with a Gaussian
  turbulent line FWHM broadening ($\Delta{V}_{\rm{turb}}$) which
  describes the local turbulence of the gas in the disk. The
  instrumental broadening ($\Delta{V}_{\rm{instr}}$) can also be
  represented by a Gaussian, which has a FWHM value of 3 km s$^{-1}$
  for CRIRES. The lowest value for $\Delta{V}_{\rm{mod}} =
  (\Delta{V}_{\rm{turb}}^2 + \Delta{V}_{\rm{instr}}^2)^{1/2}$ is
  therefore set to 3 km s$^{-1}$ and any additional broadening
  represents an increase in the value of $\Delta{V}_{\rm{turb}}$. For
  temperatures up to 1000 K, typical of the molecular gas in the
  inner disk, the thermal broadening of the CO lines of up to 1.3 km
  s$^{-1}$ is negligible.

\subsubsection{Fits with a Keplerian model with a power-law temperature profile} \label{fit} 

The sources TW Hya, VV Ser and AS 205 A were chosen as illustration
since their fundamental CO emission lines represent the three main
different types of line profiles in our large sample. Each of these sources and their fits are presented here and shown in Fig.~\ref{fig:linemodelsum}.\\

\textbf{Single peaked narrow line profile -- TW Hya:} A good fit was
found to the narrow single peaked line profile of the TW Hya
disk. This fit was achieved by using a low inclination angle of $i$ =
4.3$\degr$, $\alpha$ = 0.4, a stellar mass of 0.7 M$_{\odot}$, $R_0$ =
0.1 AU, $T_0$ = 1100 K, $R_{\rm{max}}$ = 1.5 AU and
$\Delta{V}_{\rm{mod}} =$3 km s$^{-1}$, see the left plot in
Fig.~\ref{fig:linemodelsum}. The adopted stellar mass and disk
inclination are taken from \citet{Pontoppidan08}. For this
  source, CO emission from the Keplerian disk model described above,
  convolved to the 3 km s$^{-1}$ instrumental resolution of CRIRES,
  provides a good fit to all lines.  Additional line broadening from
  turbulence is not needed to explain the line profiles. This low
turbulence is consistent with the low turbulent velocity of $\sim$0.1
km s$^{-1}$ inferred for the outer disk of TW Hya by \citet{Qi06},
although that value refers to much cooler gas
at large radii in the disk.\\

\textbf{Double peaked line profile -- VV Ser:} VV Ser is an example of a
disk with double peaked line profiles extending out to velocities of
$\pm$ 40 km s$^{-1}$. VV Ser is a Herbig Ae/Be star of spectral type A2-B6
with a mass of 2.6 $\pm$ 0.2 M$_{\odot}$ and an inclination of
65-75$^o$ \citep{Pontoppidan07}. A good fit between the model and the
data is shown in the center panel in Fig~\ref{fig:linemodelsum}.  The
estimated parameters are $M_{\star}$ = 2.6 M$_{\odot}$, $i$ =
70$\degr$, $T_0$ = 3500 K, $\alpha$ = 0.45, $R_0$ = 0.08 AU,
$R_{\rm{max}}$ = 11 AU and $\Delta{V}_{\rm{mod}} = $3 km
s$^{-1}$. However in this case the parameter space is rather large
since smaller inclination angles will give equally good fits if a
higher stellar mass is used. The main point here is that a good fit
can readily be obtained with reasonable parameters for a Keplerian
model.\\

 \textbf{Single peaked broad-based line profile -- AS 205 A:} No equally
good fit could be found for the emission lines from AS 205 A. An
approximate fit is shown in red in the right plot of
Fig.~\ref{fig:linemodelsum} with a model using an
inclination angle of $i$ = 22$\degr$, $R_{0}$ = 0.04 AU,
$R_{\rm{max}}$ = 5 AU, $ \Delta{V}_{\rm{mod}} = $3 km s$^{-1}$, $T_0$
= 1100 K, $\alpha$ = 0.3, and a mass of 1.0 M$_\odot$. The mass is the
same and the inclination angle is close to that given by
\citet[][1.0\,M$_{\odot}$ and 25$\degr$, respectively] {Andrews09}. 
As for TW Hya, the millimeter data of \citet{Andrews09} 
refer to much larger radii (typically $>$50 AU) than the CRIRES 
observations that the model is fitted to. 
A lower inclination angle of 10$\degr$ gives a better fit to the narrow
central part of the line but results in poor fitting of the outer
wings of the profile, as presented by the green profile in Fig.~\ref
{fig:linemodelsum}. The blue line represents the best fit to the line wings, which is
achieved by increasing the inclination angle to 27$\degr$ and taking
$\alpha$ to be 0.36.

\begin{figure*}[htb]
\centering
{
 \includegraphics[width=60mm]{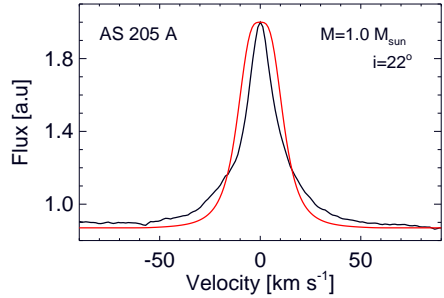}
 \caption{Fit to the broad-based single peaked line of AS 205 A
   (black) using a standard Keplerian model with a power-law temperature profile (red) including an enhanced
   line broadening parameter ($\Delta{V}_{\rm{mod}}$) of 8 km s$^{-1}$.  The other parameters are $i$ =
   22$\degr$, $R_{0}$ = 0.04 AU, $R_{\rm{max}}$ = 5 AU,
   $T_0$ = 1100 K, $\alpha$ = 0.3 and a mass of 1.0 M$_\odot$ (as in
   $\S$\ref{fit}). }
 \label{fig:as205_oneline}
}
\end{figure*}

One option to improve the fit to the AS 205 A data could be to
  increase the turbulent broadening so that the central dip is filled
  in. Indeed, increasing $\Delta V_{\rm mod}$ to 8 km s$^{-1}$,
  removes the central double peak (see
  Fig.~\ref{fig:as205_oneline}). This fit is unable to simultaneously
  match both the narrow peak and the broad base of the profile,
  however.

In summary, the fits to TW Hya and VV Ser show that both narrow single
peaked and double peaked line profiles can be reproduced with a simple
Keplerian model with reasonable parameters. However a single peaked
line with broad wings cannot be well explained with this type of
standard Keplerian model as long as the temperature gradient is
described by a continuous power-law.

\subsubsection{Keplerian disk models with a non-standard temperature profile} \label{iterative}

\begin{figure*}[htb]
\centering
{
 \includegraphics[width=100mm, angle=0.0]{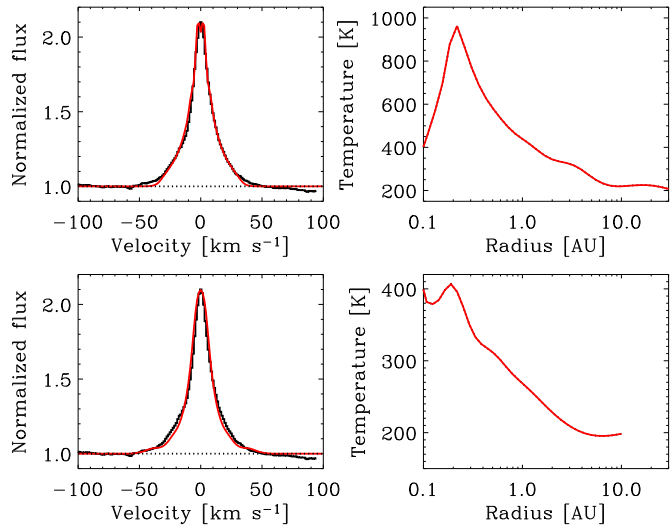}
 \caption{Left panel: Keplerian model (red) using an iterative
     inversion method to fit the velocity profile of the AS 205 A CO
     $\rm{v} = 1-0 $ P(8) line (black). Right panel: The excitation
     temperature as a function of radius needed to create the modeled
     lines. The models in the upper panels include extended emission out to 30 AU, a stellar mass of 1.4 M$_\odot$, an inclination of
     25$\degr$ and $\Delta V_{\rm mod}$=3 km s$^{-1}$. The models in
     the lower panels include emission out to 10 AU, a stellar mass of
     1.0 M$_\odot$, an inclination of 22$\degr$ and an increased $\Delta
     V_{\rm mod}$=7 km s$^{-1}$.}
 \label{fig:iterative}
}
\end{figure*}

Since disk models with a standard temperature and density
  power-law do not fit the data, one complementary approach is to investigate what physical distribution would be needed to
  reproduce the line profiles within a Keplerian model. For this, an
  iterative approach is used in which the observed line profile is
  `inverted' to determine the temperature distribution that would be
  consistent with the data. Specifically, the disk is divided into rings and the temperature of each ring is adjusted to match the line profile. The left panels of
  Fig.~\ref{fig:iterative} present the model compared with the
  velocity profile of the AS 205 CO ${\rm v}=1-0$ P(8) line whereas
  the right plots show the corresponding excitation temperature
  profiles with radius. For LTE excitation and optically thin
  emission, the latter can also be viewed as an intensity profile with
  radius. The upper plots use the same parameters as in $\S$\ref{fit}
  for AS 205 A but with a higher stellar mass of 1.4 M$_\odot$, a higher
  inclination angle of $i$=25\degr and including extended emission out
  to 30 AU. The lower plots show a model using the same parameters as
  in $\S$\ref{fit} but with an extended emission out to 10 AU and
  increasing the $\Delta V_{\rm mod}$ to 7 km s$^{-1}$. Both of these
  models fit the line profiles well, illustrating that a Keplerian
  disk model with an unusual temperature distribution can explain the
  data. The emission is much more extended, however, than in the case
  of the standard disk models with a power-law temperature profile: out to 30 AU in the case of no turbulence or out to 10 AU with increasing $\Delta V_{\rm
    mod}$. These models will be further tested in $\S$\ref{disc_rot_disk} using the constraints on the spatial extent of the observed emission found in $\S$\ref{extended_sec}.

\begin{figure*}[htb]
\centering
{
 \includegraphics[width=70mm, angle=0.0]{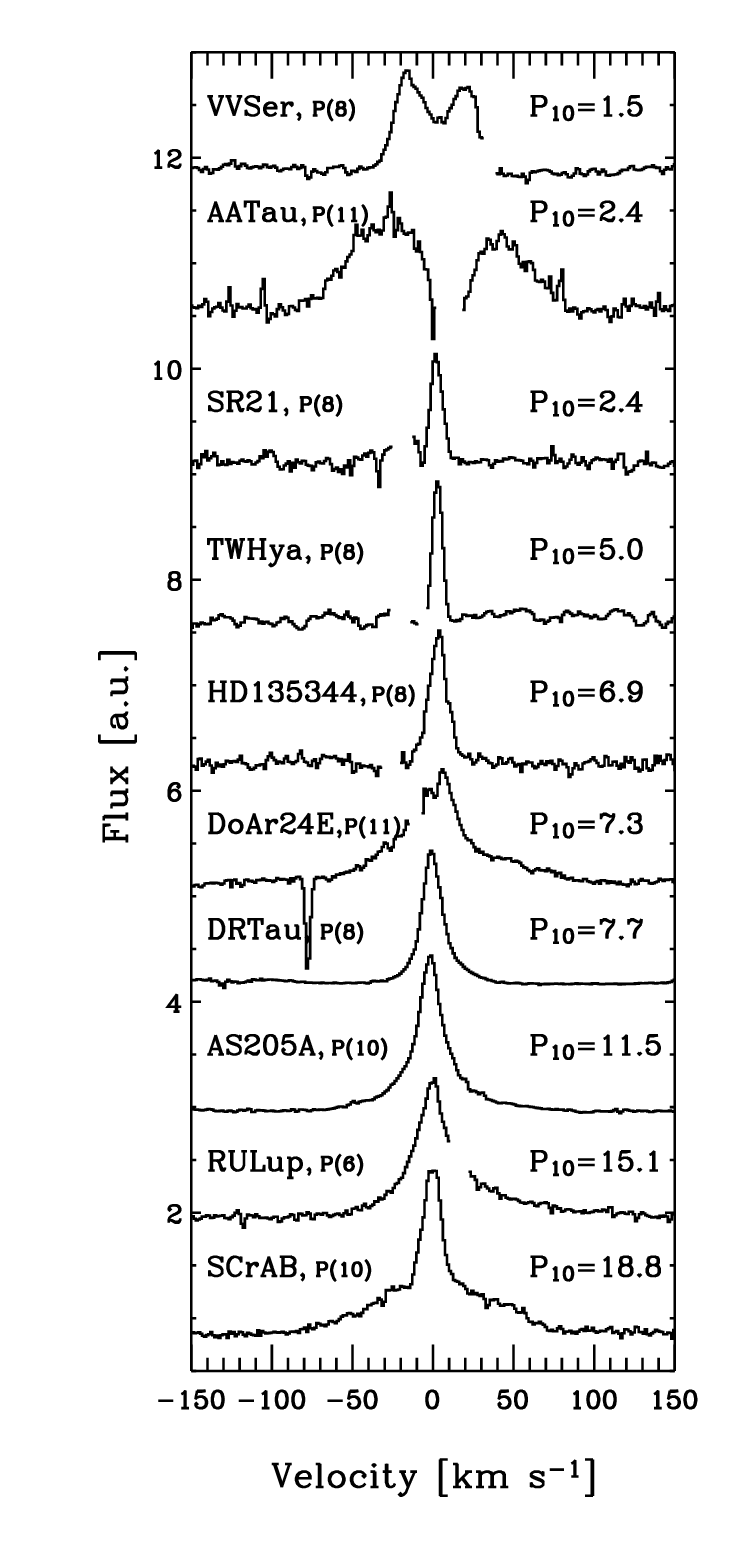}
 \caption{The line profile parameter
 $P_{10}$=$\Delta{V}_{10}$/$\Delta{V}_{90}$ is presented here for a sample
 of selected sources. Note that this parameter is clearly larger for
 the broad-based single peaked line profiles. The $P_{10}$-value is an
 average of the lower $J$-transition lines (up to P(14)). The $J$-transition corresponding to each plotted profile is given in the figure. The
 narrow absorption line seen toward DoAr24E A is due to $^{13}$CO
 absorption in the foreground cloud.}
 \label{fig:peakinessfig}
}
\end{figure*} 

\begin{figure*}[htb]
\centering
{
 \includegraphics[width=160mm]{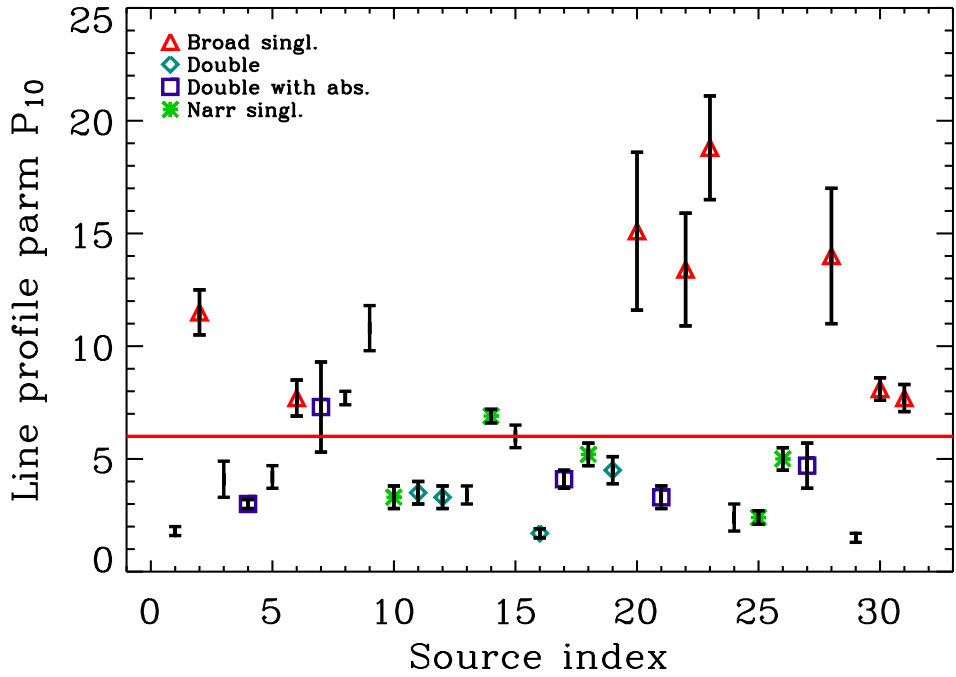}
 \caption{The line profile parameter $P_{10}$ for different
 sources. The index numbers represent the different sources, see Table
 1. Each type of line profile is given by a specific symbol. Data
 points without a symbol are for sources without clearly defined line
 profiles. The red line gives the maximum line profile parameter
 achieved using a standard Keplerian disk model with a power-law temperature profile.}
 \label{fig:line_prof}
}
\end{figure*}

\begin{table*}
\footnotesize
\caption{The peakiness parameter $P_{10}$, the accretion luminosity
and the line/continuum ($L/C $) ratio for a selected sample of protoplanetary disks.}
 \begin{minipage}[t]{\columnwidth}
  \renewcommand{\footnoterule}{} 
 \centering
  \thispagestyle{empty}
   \begin{tabular}{c l l c r c}
   \hline
    \hline 
    Index number & Source & $P_{10}$\footnote{Values refer to an average of the lower $J$-
    transitions. Uncertainties are indicated in parentheses} & $L/C$   & Acc. lum.  & Ref.\footnote{References - (1) \citet{Gullbring1998}; 
    (2) \citet{Valenti1993}; (3) \citet{Herczeg08}; (4) \citet{Natta06}; (5) \citet{Garcia06}; and (6) \citet
    {Hartmann1998}.}\\
  & & & & [log(L$_\odot$)] & \\
   \hline 
1 & AA Tau & 1.8 (0.2) & 1.5 & -1.6  & 1 \\
2 & AS 205 A & 11.5 (1.0) & 2.1 & 0.2\footnote{For both A and B component} & 2 \\
3 & CV Cha & 4.1 (0.8) & 1.3 & - & - \\
4 & CW Tau & 3.0 (0.2) & 1.4 & -1.3 & 2 \\
5 & DF Tau & 4.2 (0.5) & 1.5 & - 0.7 & 3 \\
6 & DR Tau & 7.7 (0.8) & 2.9 & - 0.1 & 2 \\
7 & DoAr24E A & 7.3 (2.0) & 1.5 & - 1.6 & 4 \\
8 & DoAr 44 & 7.7 (0.3) & 1.4 & - & - \\
9 & EX Lup & 10.8 (1.0) & 1.3 & - & - \\
10 & FN Tau & 3.3 (0.5) & 1.5 & -  & - \\
11 & GQ Lup & 3.5 (0.5) & 1.4 & - & - \\
12 & Haro 1-4 & 3.3 (0.5) & 1.4 & - & - \\
13 & Haro 1-16 & 3.4 (0.4) & 1.3 & - & - \\
14 & HD135344B & 6.9 (0.3) & 1.2 & -0.9 & 5 \\
15 & HD142527 & 6.0 (0.5) & 1.2 & 0.0 & 5 \\
16 & IRS 48 & 1.7 (0.2) & 1.2 & - & - \\
17 & IRS 51 & 4.1 (0.4) & 1.2 & - & - \\
18 & LkHa 330 & 5.2 (0.5) & 1.3 & - & - \\
19 & RNO 90 & 4.5 (0.6) & 1.5 & - & - \\
20 & RU Lup & 15.1 (3.5) & 1.8 & - 0.4 & 3 \\
21 & RY Lup & 3.3 (0.5) & 1.2 & - & - \\
22 & S CrA A & 13.4 (2.5) & 2.0 & - & - \\
23 & S CrA B & 18.8 (2.3) & 1.8 & - & - \\
24 & SR 9 & 2.4 (0.6) & 1.1 & -1.4 & 4 \\
25 & SR 21 & 2.4 (0.3) & 1.2 & $<$-1.9 & 4 \\
26 & TW Hya & 5.0 (0.5) & 2.3 & -1.4 & 3 \\
27 & VSSG1 & 4.7 (1.0) & 1.5 & - 0.4 & 4 \\
28 & VV CrA A & 14.0 (3.0) & 1.4 & - & - \\
29 & VV Ser & 1.5 (0.2) & 1.1 & 1.2 & 5 \\
30 & VW Cha & 8.1 (0.5) & 2.0 & - 0.2 & 6 \\
31 & VZ Cha & 7.7 (0.6) & 2.2 & - 1.1 & 6 \\
\hline	
\end{tabular}
\end{minipage}
\label{tab:acc_peak}
\end{table*}

\subsection{Line profile parameter} \label{peakiness}

A parameter is defined to quantify the difference in line
  profiles between the broad single peaked sources (category C) and
  the other sources in the sample. This so-called line profile
parameter $P_{10}$ describes the degree to which line profiles have a
broad base relative to their peak. The line profile parameter $P_{10}$
is therefore defined as the full width ($\Delta{V}_{10}$) of
the line at 10\% of its height divided by the full width at
90\% ($\Delta{V}_{90}$) of its height, $P_{10}$ =
$\Delta{V}_{10}/\Delta{V}_{90}$. The broad-based single peaked lines
(category C) have a higher value of the line profile parameter
relative to the double and narrow single peaked lines, see
Fig.~\ref{fig:peakinessfig}.

\begin{figure*}[htb]
\centering
{
 \includegraphics[width=160mm]{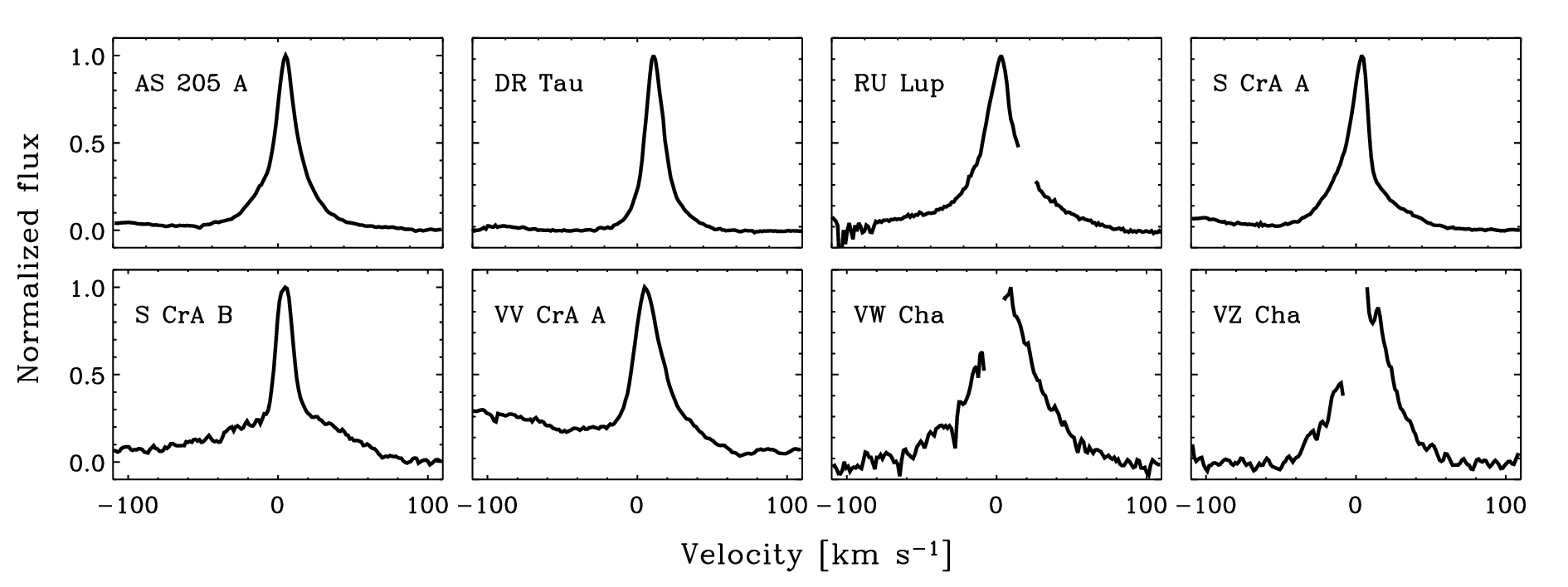}
 \caption{The P(8) $^{12}$CO line is plotted for the entire sample broad-based single peaked sources, apart from VW Cha for which the P(10) line is
 presented because the P(8) line is affected by a strong telluric
 feature at the time of observation. The lines have been continuum
 subtracted and normalized. The asymmetry of the VV CrA A line is
 caused by line blending by the CO $\rm{v} = 3-2$ R(5) line. This line
 overlap adds intensity on the blue side of the $^{12}$CO P(8) line. }
 \label{fig:singlepeak}
}
\end{figure*}

A summary of the line profile parameter values for 31 selected T Tauri
stars from the total sample of $\sim$50 T Tauri stars with CO emission
is presented in Fig.~\ref{fig:line_prof} and Table
\ref{tab:acc_peak}. These 31 T Tauri stars were selected because their
CO emission line profiles have high $S/N$ and are not contaminated 
by strong telluric or absorption lines. Figure
\ref{fig:line_prof} shows that the double peaked (turquoise diamonds)
and narrow single peaked (green stars) sources all have a line
profile parameter of $<$6. 
For reference, a Gaussian line profile has a $P_{10}$-value of
4.7. The selected sample of 8 broad centrally peaked sources (red
triangles) all have a $P_{10}$-value of $>$6. An overview of the
normalized and continuum-subtracted $^{12}$CO P(8) line profiles of
this subset is presented in Fig.~\ref{fig:singlepeak}. These lines are
typically symmetric around line center and have a narrow top and a
very broad base that extends out to $\pm$100 km s$^{-1}$ in several
cases. Four other sources, DoAr24E A, DoAr44, EX Lup and HD135344B, also
have higher ($>$6) $P_{10}$- values but are not selected for our sample (see
$\S$\ref{source_selection} for source selection criteria).

\subsection{Model line profile parameters}

\begin{figure*}[htb]
\centering
{
 \includegraphics[width=160mm]{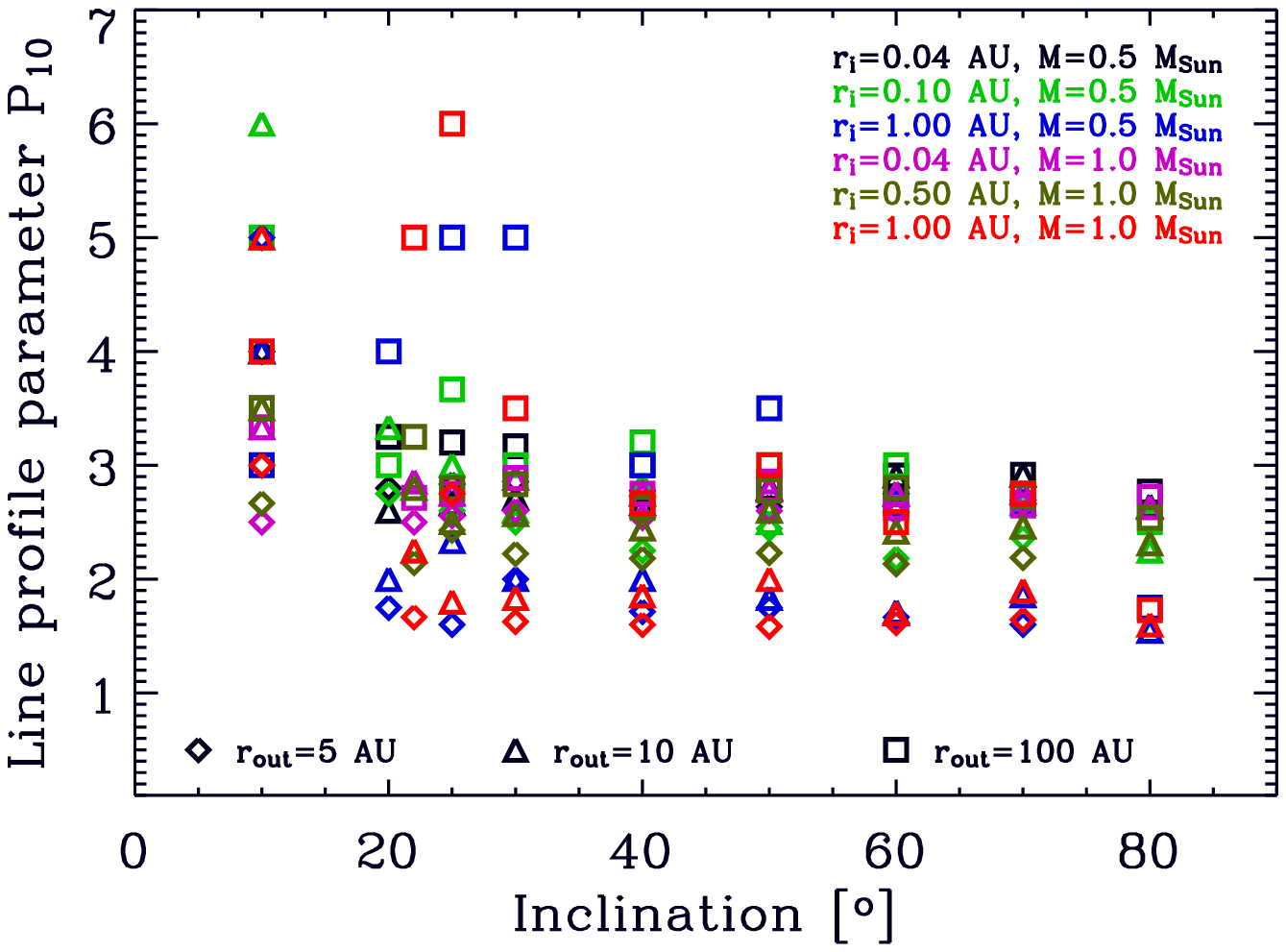}
 \caption{Line profile parameter $P_{10}$ versus inclination for a
 large model grid of standard Keplerian models with a power-law temperature profile. Note that $P_{10}$ does not become larger than 6 and shows no trend with inclination for $i>20^o$.}
 \label{fig:modelgrid}
}
\end{figure*}

To derive the maximum line profile parameter consistent with a
standard Keplerian model with a power-law temperature profile, a grid of Keplerian line profiles was constructed. Three parameters in the model were set to fixed values
that give strongly peaked line profiles; $T_0$ = 1100 K, $\alpha$ =
0.30 and $\Delta{V}_{\rm{mod}}$ = 3 km s$^{-1}$. The following
parameters were varied between $M_{\star}$ = 0.5 -- 1.0 M$_{\odot}$,
$i$ = 10 -- 80$\degr$, $R_0$ = 0.04 -- 1.0 AU and $R_{\rm{max}}$ = 5
-- 100 AU and the $P_{10}$- value was calculated for each produced
line profile. An overview of the $P_{10}$-values versus inclination is
shown in Fig.~\ref{fig:modelgrid}. It is found that a typical standard
Keplerian model with a power-law temperature profile can achieve a maximum $P_{10}$-value of about 6, with
the highest values found at low inclination. This conclusion is
unchanged if $\Delta V_{\rm mod}$ is increased to values as large as
$\sim$10 km s$^{-1}$.  The upper limit of $P_{10} \approx$ 6 that the
Keplerian model can reproduce is presented as a red line in
Fig.~\ref{fig:line_prof}. Figure \ref{fig:modelgrid} also indicates
that the inclination of the source, invoked to explain the different
widths of Herbig Ae disk profiles \citep{Blake04}, should not affect
the value of $P_{10}$ significantly except at very low inclinations,
since it enters both $\Delta{V}_{10}$ and $\Delta{V}_{90}$.

Note that the Keplerian models always require a low
  inclination angle of 20 -- 30$\degr$ to obtain high values of the
  line profile parameter $P_{10}$, regardless of whether or not a
  large turbulent broadening is included. If inclination angles within 5$\degr$ of the range $i$ = 20 -- 30$\degr$ are considered
  necessary for the broad single peaked line profiles, then 6 $\pm$ 3
  sources of the total sample of 50 T Tauri stars would have broad,
  single-peaked line profiles, which is consistent with our subsample
  of 8 sources.  Millimeter interferometry data for AS 205 A and DR
  Tau yield inclinations of 25$\degr$ and 37$\pm$ 3$\degr$,
  respectively \citep{Andrews_thesis,Andrews09,Isella09}, which are
  roughly consistent with the model requirements.

\subsection{$P_{10}$-value versus the line-to-continuum ratio and source selection.}\label{source_selection}

\begin{figure*}[htb]
\centering
{
 \includegraphics[width=140mm]{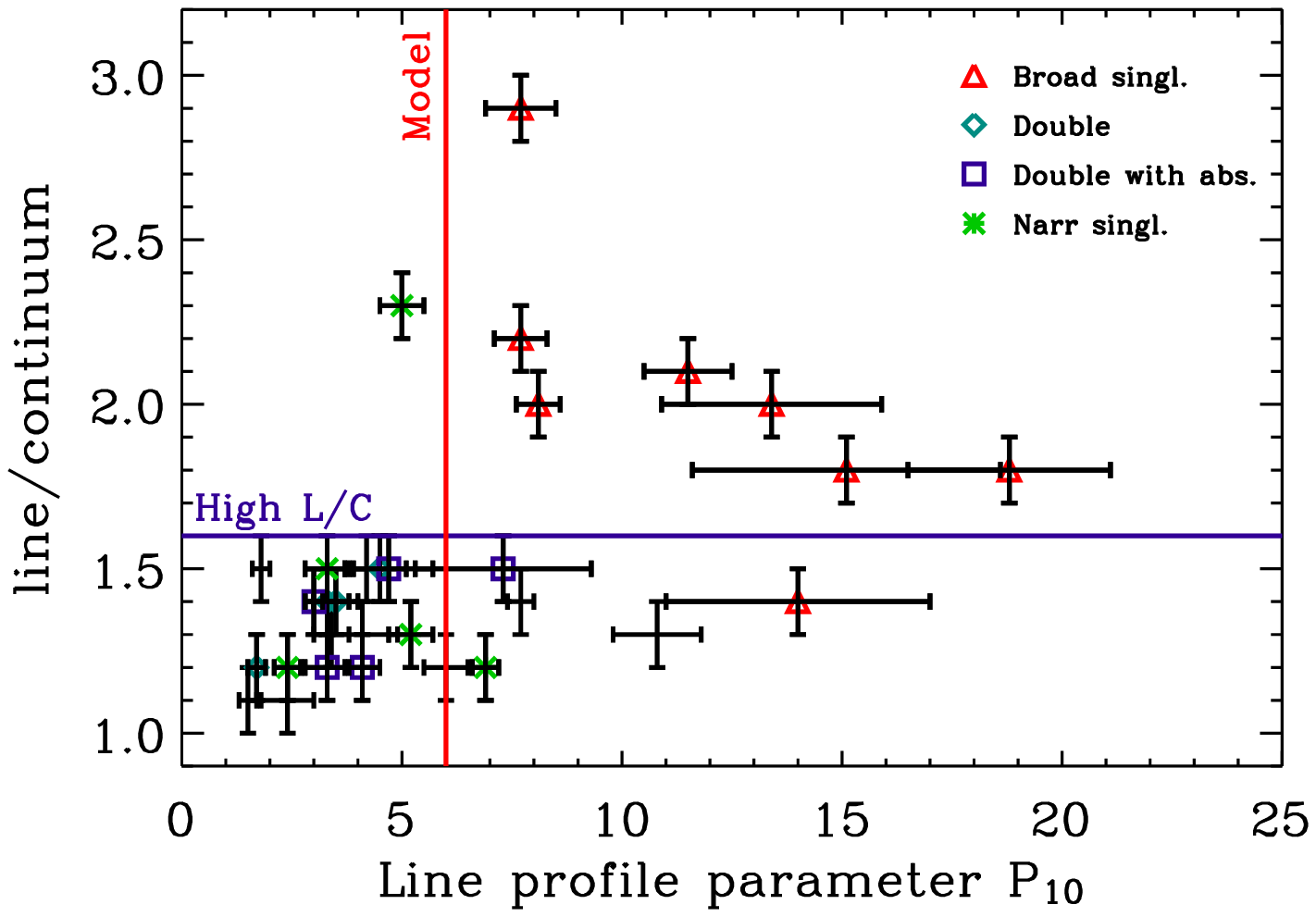}
 \caption{The line to continuum ratio relative to the line profile
  parameter $P_{10}$. Each type of line profile is given by a specific
  symbol. Data points without a symbol do not have a clearly defined
  line profile. The broad single peaked sample includes sources
  with either a line/continuum ratio $>$ 1.6 (blue line) and $P_{10}>6$ (red
  line) or solely $P_{10}>10$.}
 \label{fig:line_cont}
}
\end{figure*}

The broad-based single peaked sources are also noteworthy for having high line-to-continuum ($L/C$) ratios in many lines, including the resolved CO ro-vibrational lines discussed here and H$_2$O lines in \citet{Salyk08}. Here, the $L/C$-ratio is defined as the ratio
between the peak line flux relative to the continuum flux. The $L/C$
ratios of the $^{12}$CO lines have been plotted against the
$P_{10}$-values in Fig.~\ref{fig:line_cont}. Typical uncertainties of
the line-to-continuum ratio are $\pm$0.1. Seven of the sources stand
out clearly with high line profile parameters and high
line-to-continuum ratios. Figure~\ref{fig:line_cont} is used to set
the criteria ($P_{10}$ $>$ 6 and $L/C$ $>$ 1.6) for the selection of
the broad-based single peaked line profiles (category C). The lower
limit for the $P_{10}$-value ($P_{10}$ $>$ 6) for the sample is set to
match the maximum line profile parameter for a line that the standard
Keplerian model with a power-law temperature profile can produce (\S 3.4). The line-to-continuum constraint
of $L/C$ $>$ 1.6 is rather arbitrary, based on the observation that
the majority of the sources have a lower value. Two other sources, EX
Lup and VV CrA A, have $L/C$ ratios less than 1.6 but have such a
high line profile parameter ($>$10) that they are considered to belong
to the category C profile sample. However, EX Lup is a variable source
that underwent an outburst in 2008. The variation of the CO line
profiles from EX Lup during and after the outburst are discussed in
Goto et al.\ (subm.). DoAr 44 and DoAr24E A are borderline cases and are
not included here because their profiles are contaminated by
absorption, limiting the analysis. HD135344B is also excluded from the
sample due to the combination of having both a low $L/C$ ratio and a
$P_{10}$-value close to 6. In addition HD135344B is a well-known
face-on disk whose profile has been well fitted with a Keplerian model
\citep{Pontoppidan08,Brown09,Grady09}. Line profiles emitted by disks
with a low inclination angle can have a line profile parameter close
to 6, as shown in Fig.~\ref{fig:modelgrid}. These selection criteria
yield in total 8 sources that are included in the broad-based single
peaked sample; these sources are marked with red triangles in
Fig.~\ref{fig:line_cont}.

\section{Characteristics for the sources with broad single peaked lines} \label{char}

In \S 3, we showed that our selected sources with broad single peaked
line profiles (category C) cannot be reproduced well with a standard
Keplerian model with a power-law temperature profile. In the following sections, we analyze the
observational characteristics of the emission to obtain further
constraints on the origin.

\subsection{Line profiles of CO isotopologues and the $\rm{v} = 2-1$ CO lines.}

 \begin{table*}
\footnotesize
\caption{List of CO isotopologues and the $^{12}$CO $\rm{v} = 2-1$, $\rm{v} = 3-2$ and $\rm{v} = 4-3$ lines.}
 \begin{minipage}[t]{\columnwidth}
  \renewcommand{\footnoterule}{} 
 \centering
  \thispagestyle{empty}
  \begin{tabular}{l c c c c c c c c c}
   \hline
    \hline
Source\footnote{x = detection in emission, a = detection in absorption and - = no detection.} & $^{12}$CO & $^{13}$CO & CO $\rm{v} = 2-1$ & CO $\rm{v} = 3-2$ & CO $\rm{v} = 4-3$ & C$^{18}$O \\
   \hline                     
AS 205 A & x & x & x & x & x & x \\                                        
DR Tau & x & x & x & x & x & x \\               
RU Lup & x & x & x & x & - & - \\    
S CrA A & x, a & x & x & x & x & x \\ 	
S CrA B & x, a & x & x & x & - & - \\     	                   
VV CrA A & x, a & x, a & x & x & x & a \\  
VW Cha & x & a & x & - & - & - \\   
VZ Cha & x & - & x & x & - & - \\           
\hline
\end{tabular}
 \end{minipage}
 \label{tab:isotopologues}
\end{table*}

\begin{figure*}[htb]
\centering
{
 \includegraphics[width=160mm]{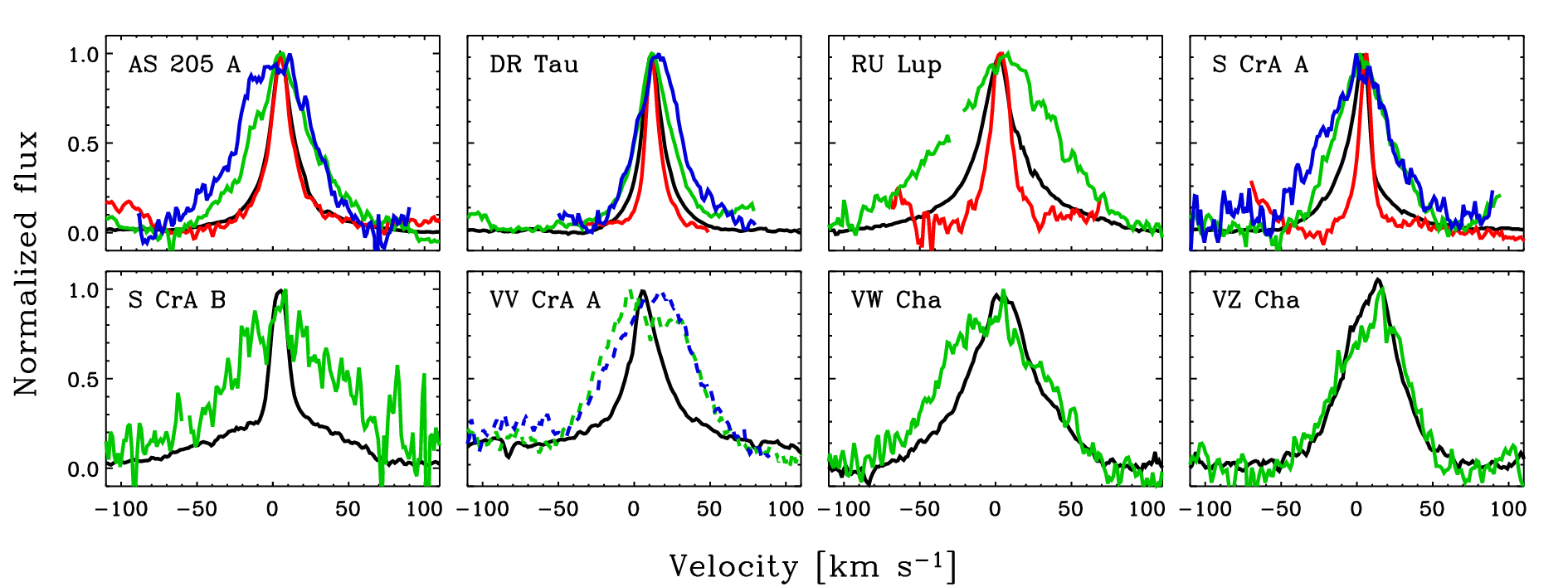}
 \caption{$^{13}$CO (red), CO $\rm{v} = 2-1$ (green) and $\rm{v} =
   3-2$ (blue) line are plotted on top of the CO $\rm{v} = 1-0$ lines
   (black) for the broad-based single peaked CO emitting sample of T Tauri
   stars. Dashed lines indicate that the line profile is constructed
   by stacked lines that all are to some degree affected by line
   overlap. All profles have been normalized to a peak flux of unity.}
 \label{fig:stack}
}
\end{figure*}

Table \ref{tab:isotopologues} summarizes the lines seen in the spectra
of the broad single peaked sources, including if they are detected in
absorption, emission or both. In the richest spectrum, that of AS 205
A, lines of $^{13}$CO, C$^{18}$O, CO $\rm{v} = 1-0$, $\rm{v} = 2-1$, $\rm{v}= 3-2$ and $\rm{v} = 4-3$ are all detected. All
of the sources, besides VZ Cha, have detections of $^{13}$CO. Four sources have a C$^{18}$O detection in emission. In addition
every source in the sample has detections of CO $\rm{v} = 2-1$, 7 of 8
sources of CO $\rm{v} = 3-2$ and 4 of 8 sources of CO $\rm{v} = 4-3$.

Fig.~\ref{fig:stack} compares the stacked, normalized and
continuum-subtracted line profiles for $^{13}$CO $\rm v=1-0$,
$^{12}$CO $\rm v=1-0$, $^{12}$CO $\rm v=2-1$ and  $^{12}$CO $\rm
v=3-2$. All of the isotopologues and the higher ro-vibrational
transitions also show a broad-based single peaked line
profile. However, the profiles of the $^{12}$CO and $^{13}$CO $\rm
v=1-0$ lines match exactly for only one source, AS 205 A.  For the
other sources, DR Tau, RU Lup, and S CrA A, the $^{13}$CO lines are
narrower than the $^{12}$CO $\rm v=1-0$ lines. 

The width of the CO $\rm{v} = 1-0$ lines at 10\% of their height (see
Table \ref{tab:peakiness_values}) are in general narrower than the CO
$\rm{v} = 2-1$ and $\rm{v} = 3-2$ lines by about 60-80\% which can be
seen in Fig.~\ref{fig:stack}. \citet{Najita03} also found a similar
trend of broader higher vibrational lines in their data. The
difference in width between the lower and higher vibrational lines
in our sample may reflect a physical difference in the location of the
emitting gas. The high fraction of CO $\rm{v} = 2-1$ and $3-2$ detections in our sample of broad-based single peaked line sources is consistent with the
\citet{Najita03} sample, where 9 of their 12 CO sources have
detections of $\rm{v} = 2-1$ emission and 3 out of 12 sources have
$\rm{v} = 3-2$ detections. However it is difficult to categorize their
line profiles due to the lower spectral resolution.

\subsubsection{Two component fits} \label{two_comp}

\begin{figure*}[htb]
\centering
{
 \includegraphics[width=120mm, angle=0.0]{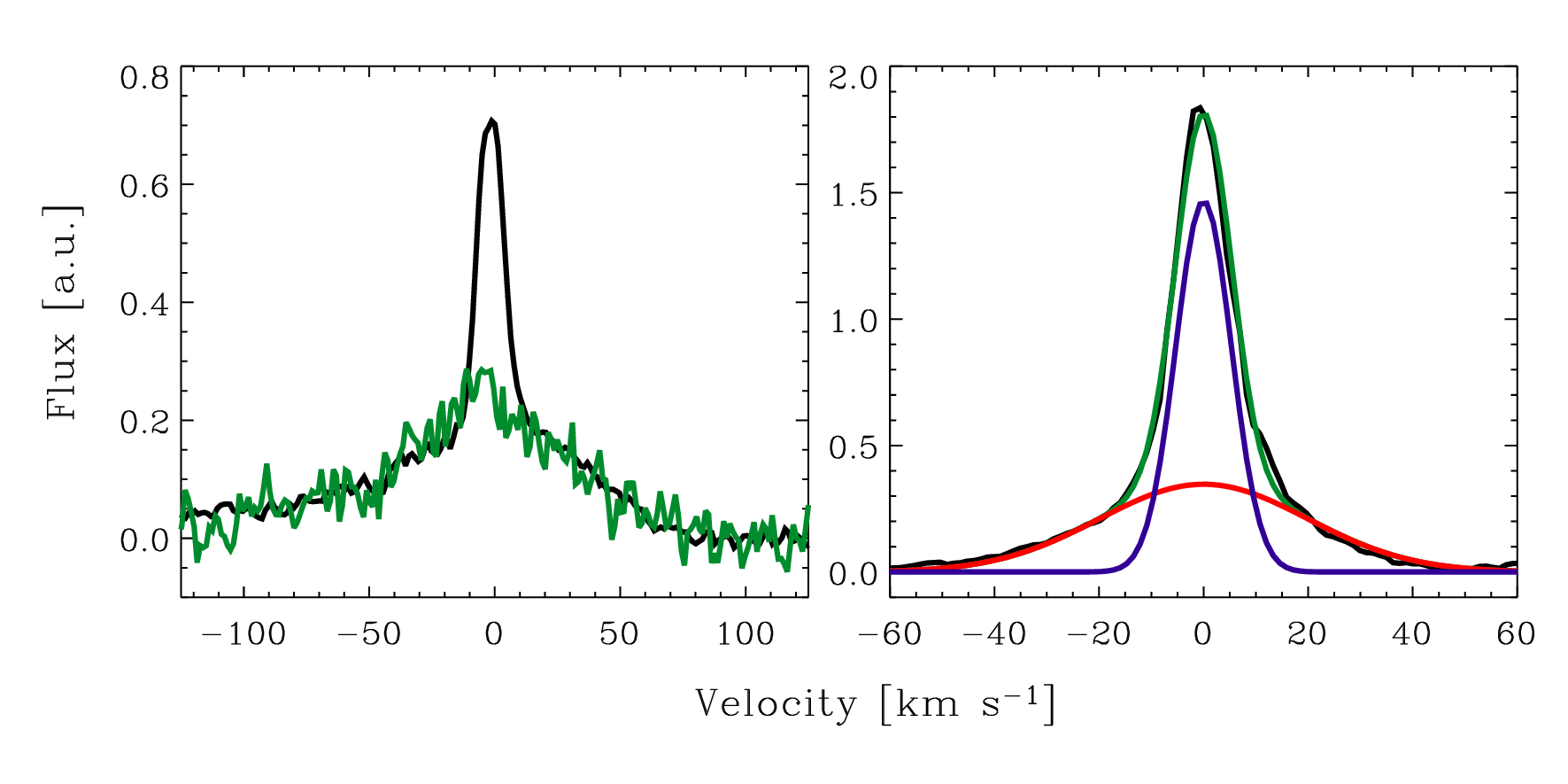}
 \caption{Left: The higher $J$-transition $^{12}$CO P(32) (green) is plotted on top of the lower $J$-transition $^{12}$CO P(8) (black) for S CrA B. The narrow component of the $^{12}$CO P(8) line decreases with increasing excitation. Right: Two Gaussian fitted lines (in red and blue) to the CO $\rm{v} = 1-0$ P(7) line of DR Tau (black). The green line represents the sum of the two Gaussian fits.}
 \label{fig:drtau_scra0}
}
\end{figure*}

  \begin{table*}
\footnotesize
\caption{The FWHM of the broad and narrow components and their average heliocentric velocities $V_{B}$(CO) and $V_{N}$(CO) relative to literature values of the heliocentric velocities for the stars $V_{h}$(star).}
 \begin{minipage}[t]{\columnwidth}
  \renewcommand{\footnoterule}{} 
 \centering
  \thispagestyle{empty}
   \begin{tabular}{l l l l l l l}
   \hline
   \hline
   Source & FWHM$_{B}$ & FWHM$_{N}$ & $V_{B}$(CO) & $V_{N}$(CO) & $V_{h}$(star)  & Ref.\footnote{References. - (1)  \citet{Melo03}; (2)  \citet{Guenther07} and (3) \citet{Ardila02}.}\\
    & km s$^{-1}$ & km s$^{-1}$ & km s$^{-1}$ & km s$^{-1}$ & km s$^{-1}$ & \\ 
   \hline
AS 205 A & 61.3 (16.3) & 14.8 (1.7)  & -4.4 (4.8) & -6.2 (0.3) & -9.4 (1.5) & 1 \\
DR Tau & 39.8 (6.3) & 13.1 (0.9) & 26.9 (1.3) & 24.4 (0.7) & 27.6 (2.0) & 3 \\
RU Lup & 96.8 (11.0) & 24.0 (3.9) & 0.4 (7.0) & -3.9 (1.1) & -0.9 (1.2) & 1 \\
S CrA A & 50.2 (6.7) & 10.7 (1.2) & -9.2 (1.4) & -5.0 (0.7) & 0.9 (0.9) & 2 \\
S CrA B & 97.6 (17.7) & 12.3 (0.5) & -4.7 (3.5) & -3.5 (0.3) & 0.9 (0.9)\footnote{Taken to be the same as S CrA A since in the literature only a value for S CrA is given} & 2 \\
VV CrA A\footnote{Broad component could not be determined due to large amount of line overlap.} & - & 19.3 (2.0) & - & -0.9 (0.2) & - & - \\
VW Cha & 70.6 (3.1) & 26.6 (2.2) & 15.4 (1.4) & 15.1 (1.9) & 17. 2 (2.0) & 2 \\
VZ Cha\footnote{Optimal fits with one Gaussian curve, results presented in the broad component columns.} & 45.1 (2.6) & - & 19.1 (1.7) & - & 16.3 (0.6) & 2 \\ 
  \hline
  \end{tabular}
 \end{minipage}
\label{tab:narrow_broad}
\end{table*}

The line profiles shown in Fig.~\ref{fig:singlepeak} are generally not
well fit by a single Gaussian profile. A good example of this is the
CO emission line profiles of S CrA B, which clearly consist of two
components. An interesting aspect is that the narrow component stands
out very prominently in the low $J$-transitions but progressively
decreases in intensity with higher $J$-transitions relative to the
broad component (Fig.~\ref{fig:drtau_scra0}). The simplest explanation
is that the components arise from different locations where the narrow
component has a lower rotational temperature than the broad component.
This is not as clearly seen for the other 7 sources in the sample.
However, \citet{Najita03} see a similar phenomenon in the broad-based
single peaked source GW Ori.

The line profiles from S CrA A and GW Ori support a hypothesis in
which two physical components combine to form broad centrally peaked
line profiles. We therefore fit the line profiles of all sources with
two Gaussians to investigate the possibility that the broad-based
centrally peaked line profiles consist of two physical components. In
the fits, all 6 Gaussian parameters (the two widths, amplitudes and central
wavelengths) are left free. For example,
Fig.~\ref{fig:drtau_scra0} shows that the CO $\rm{v} = 1-0$ P(7) line from
DR Tau is well-fit with two Gaussian profiles, one narrow and one
broad.

 \begin{figure*}[htb]
\centering
{
 \includegraphics[width=160mm]{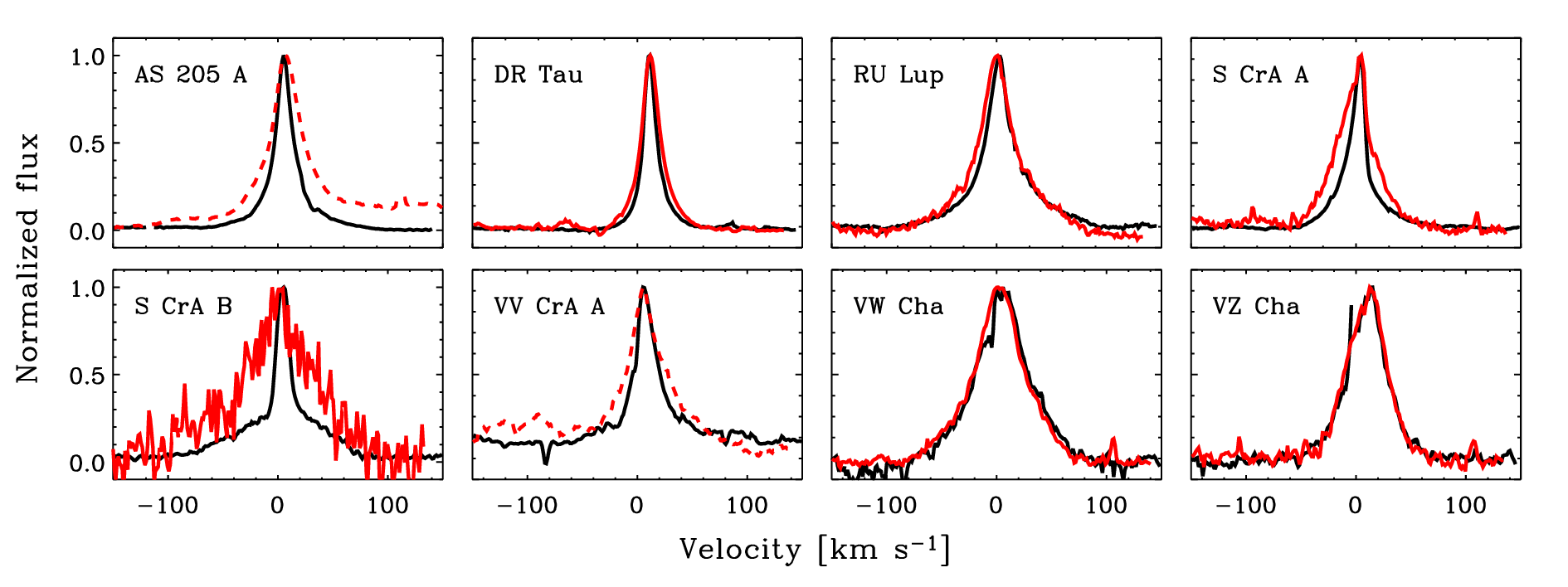}
 \caption{Comparison of the lower $J$ and higher $J$ profiles.  The
   red line represents the stacked higher $J$-transitions ($J$ =
   P(22)-P(32)) and the black line the lower $J$-transitions (up to
   $J$ = P(14)) of $\rm{v} = 1-0$ CO. The baseline for the higher
   stacked $J$-transitions for AS 205 A and VV CrA A (dashed red) is not
   straight because of line overlap and blend. All profiles have been
   normalized to a peak flux of unity.}
 \label{fig:high_low}
}
\end{figure*}

  \begin{table*}
\footnotesize
\caption{The averaged line profile parameter $P_{10}$ for high and low $J$-transitions for the sample of sources with single peaked CO $\rm{v} = 1-0$ emission lines. $\Delta{V_{10}}$ is the averaged width of the line at 10\% of its height. }
 \begin{minipage}[t]{\columnwidth}
  \renewcommand{\footnoterule}{} 
 \centering
  \thispagestyle{empty}
  \begin{tabular}{l l l l}
   \hline
   \hline
    &  \multicolumn{2}{c}{$P_{10}$} \\
   \cline{2-3}
  Source & (low $J$) & (high $J$) & $\Delta{V_{10}}$ [km s$^{-1}$] \\
   \hline       
AS 205 A & 11.5 (0.5)\footnote{Value in parentheses indicates the spread in $P_{10}$ for different $J$. Low-$J$ includes selected lines up to P(14). High-$J$ includes lines from P(22) to P(32).} & - & 62.8 (3.9) \\                                           
DR Tau & 7.7 (0.8) & 6.6 (0.6) & 40.2 (1.0) \\             
RU Lup & 15.1 (3.5) & 10.5 (2.3) & 109 (11.1) \\  	
S CrA A & 13.4 (2.5) & 7.9 (0.2) & 56.9 (3.4) \\     	              
S CrA B & 18.8 (2.3) & 12.5 (1.0) & 102.6 (6.9) \\ 
VW Cha & 8.1 (0.5) & 8.4 (0.3) & 115.0 (7.0) \\    
VZ Cha & 7.7 (0.6) & 7.7 (0.9) & 79.2 (2.4) \\     
  \hline
\end{tabular}
 \end{minipage}
\label{tab:peakiness_values}
\end{table*}

Two Gaussian profiles are needed for 7 of the 8 broad-based single
peaked sources to find an optimal fit to their line profiles. The
resulting FWHM$_{B}$, FWHM$_{N}$, $V_{B}$(CO) and $V_{N}$(CO) are
presented in Table \ref{tab:narrow_broad}. These results show that the
broad-based single peaked line profiles (category C) can be fitted
using a narrow component with a FWHM varying between $\sim$ 10 - 26 km
s$^{-1}$ and a broad component with FWHM of $\sim$ 40 - 100 km
s$^{-1}$. The central velocities of the two components are generally
the same within the errors. Only the line profile of VZ Cha was
adequately fit with a single Gaussian.

The results presented in Table \ref{tab:narrow_broad} are based on
fits to the CO $\rm{v} = 1-0$ lines that are chosen to be as
uncontaminated from line overlap as possible. The uncertainties in
Table \ref{tab:narrow_broad} are the standard deviations of the fit
parameters of individually fitted lines. The spread within a parameter
for a given source is caused by uncertainties in the Gaussian fits, by
overlapping weaker lines that may affect the width and position of the
broad component, by errors of around $\sim$1.0 km s$^{-1}$ in the
wavelength calibration (which can vary between different detectors,
see $\S$ \ref{reduction}), and degeneracies of the fits between the
broad and the narrow component. The large amount of line overlap in VV
CrA A makes measurements of the broad component highly unreliable so
the fits are not presented here.

Table \ref{tab:peakiness_values} and Fig.~\ref{fig:high_low} show that
the relative width of the peak and base as represented in $P_{10}$ is
approximately constant with increasing $J$ for most of the selected
category C sources, with the exceptions of S CrA A and B. For these
two sources the $P_{10}$-value decreases with increasing $J$ (Table
\ref{tab:peakiness_values}) and the line profiles become dominated by
the broad wings at the higher $J$-transitions, see
Fig.~\ref{fig:drtau_scra0}. AS 205 A and VV CrA A are excluded in the
calculations of the line profile parameter because of their large
amount of overlapping lines that cause difficulties in reliably
measuring the 10\% and 90\% line widths (for AS 205 A, just in the
higher $J$-transitions, see Fig.~\ref{fig:high_low}). The lack of
variation in $P_{10}$ towards higher $J$-transitions for all sources
besides S CrA A and B, combined with the absence of a velocity
  shift, suggest that the
  broad and narrow components may be formed in related
  physical regions.

\subsubsection{Central velocities}

\begin{figure*}[htb]
\centering
{
 \includegraphics[width=90mm]{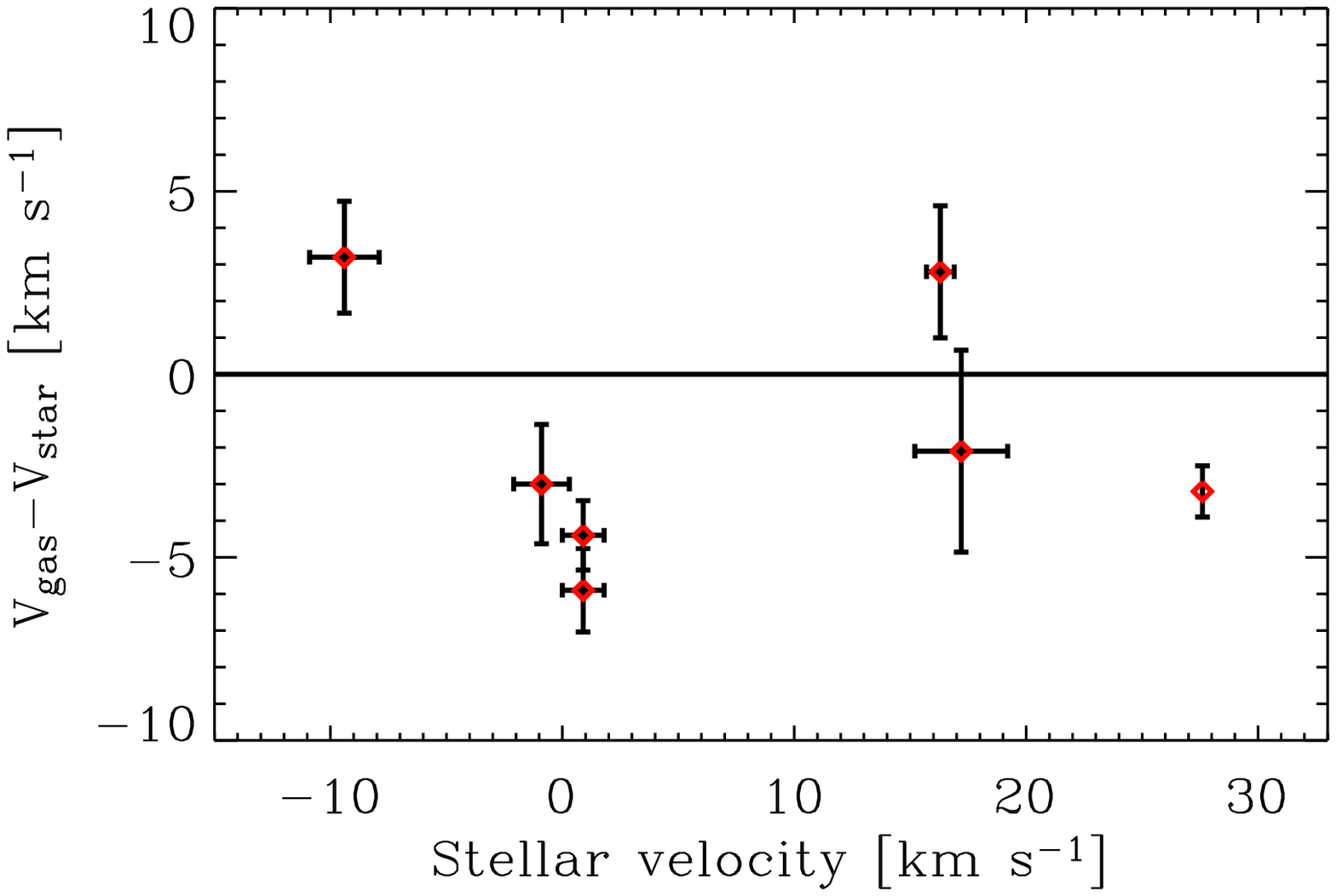}
 \caption{Velocity shifts $V_{\rm{gas}}$ - $V_{\rm{star}}$ as functions of $V_{\rm{star}}$. The gas velocity refers to the central velocity of the narrow component of the CO $\rm{v} = 1-0$ lines.}
 \label{fig:vel_diff}
}
\end{figure*}

The central velocities of the CO gas and stellar photosphere are
compared to look for differences indicative of a wind or outflow,
which could explain the narrow low velocity peak. The stellar radial
velocities in Table \ref{tab:narrow_broad} are taken from literature
measurements of photospheric optical absorption lines.  The
photospheric absorption lines for our sample of sources are generally
heavily veiled by continuum emission related to accretion, leading to
uncertainties in the stellar velocity measurements of the order of 1-2
km s$^{-1}$.

The differences between the molecular gas radial velocity of the
narrow component and the stellar radial velocity are small with shifts
of $|V_{\rm{CO}}$ - $V_{\rm{star}}|$ $\lesssim$ 5 km
s$^{-1}$, see Fig.~\ref{fig:vel_diff}. The shifts are more commonly
seen towards the blue rather than the red, indicating that the CO gas
is moving towards us relative to the star. However, the velocity shifts
are only 1-- 2$\sigma$ and may not be significant.

The small velocity differences rule out an origin of the lines in a
fast moving disk wind or outflow. However it is still possible that
the lines are emitted by gas in a slow disk wind. \citet{Najita03}
also do not see any significant velocity shifts between the radial
velocity of the star and the gas within their few km s$^{-1}$
uncertainties.

\subsection{Excitation temperatures}

\subsubsection{Rotational temperatures}\label{exc_comp} 

The extracted fluxes of the $^{12}$CO $\rm{v} = 1-0$ and $^{13}$CO
lines are used to produce excitation diagrams and thus calculate the gas
rotational temperatures. This calculation is done by using the
Boltzmann distribution, which assumes that the excitation in the gas
can be characterized by a single excitation temperature and that the
emission is optically thin;

\begin{equation}
\label{eq:bol}
\frac{N_i}{g_i} = \frac{N_\nu}{Q_r(T_{\rm{rot}})} e^{-\frac{E_i}{kT_{\rm{rot}}}}
\end{equation}

and where the column density $N_{i}$ of the upper level $i$ is described by,

 \begin{equation}
\label{eq:Ni}
N_i= \frac{4\pi F_i}{\Omega A_{ij}h\nu} 
\end{equation} 

where the solid angle $\Omega$ = $D_{\rm em}$/${d^2}$ includes the emitting area $D_{\rm em}$ and the distance $d$ to
the object. The other quantities are the Einstein coefficient $A_{ij}$
[s$^{-1}$], the frequency $\nu$ [cm$^{-1}$] and the measured line flux
$F_i$ [W m$^{-2}$], the statistical weight $g_i$ of the upper level,
the total column density $N_\nu$ for vibrational level $\rm{v}$, the
excitation temperature $T_{\rm{ex}}$ and the rotational partition
function $Q_r(T_{\rm{ex}})$. Plotting the upper energy levels $E_{i}$
against $\ell$n$(F_i$/$A_{ij}\nu{g}_{i})$ will give the
rotational temperature $T_{\rm{rot}}$ by measuring the slope of the
fitted line given by -1/$T_{\rm{rot}}$.

\begin{figure*}[htb]
\centering
{
  \includegraphics[width=140mm, angle=0]{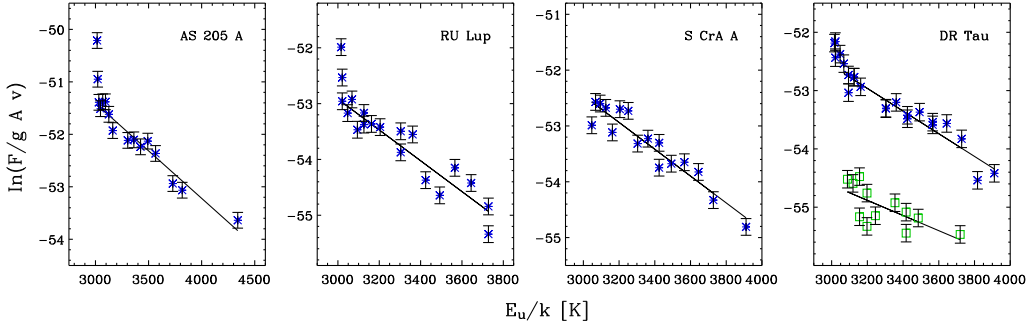}
 \caption{Rotational diagrams for T-Tauri stars AS 205 A, DR Tau, RU Lup and S CrA A. Blue stars represent $^{13}$CO lines and green squares C$^{18}$O lines.}
 \label{fig:13COrot}
}
\end{figure*}

\begin{table*}
\footnotesize
\caption{Rotational temperatures $T_{\rm rot}$ derived from the $^{13}$CO lines, together with the vibrational temperatures $T_{\rm{vib}}$.}
 \begin{minipage}[t]{\columnwidth}
  \renewcommand{\footnoterule}{} 
 \centering
  \thispagestyle{empty}
  \begin{tabular}{l c c}
   \hline
    \hline 
   Source & $T_{\rm rot}$ [K] & $T_{\rm{vib}} [K]$ \\
   \hline
AS 205 A & 550 $\pm$ 40 & 1740 $\pm$ 150 \\
DR Tau & 510 $\pm$ 40 & 1680  $\pm$ 140 \\
RU Lup & 360 $\pm$ 20 & - \\
S CrA A & 420 $\pm$ 30 & 1730 $\pm$ 140 \\
  \hline
\end{tabular}
\end{minipage}
\label{tab:rottemp}
\end{table*}

The rotation diagrams for the $^{12}$CO lines are not linear, which
suggests that the lines are optically thick. In addition, UV-pumping
can cause non-linearity between the different transitions
\citep{Krotkov1980}. To overcome this problem, rotational temperatures
have been estimated using only the $^{13}$CO emission lines for the 4
sources, AS 205 A, DR Tau, RU Lup and S CrA A, for which sufficiently
high $S/N$ emission lines are detected, see Fig.~\ref{fig:13COrot} and
Table \ref{tab:rottemp}. The relative error of the line fluxes is
taken to be 15\% based on the noise in the spectra and uncertainties
in the baseline fits. Only lines with $J_u > $1 are taken into
  account in the fit.
  
While $^{13}$CO is significantly less abundant than $^{12}$CO,
$^{13}$CO lines may also be optically thick. $^{13}$CO lines with
$\tau\approx$1 have, for example, been seen in the disk around Herbig
star AB Aur by \citet{Blake04}. The inferred rotation temperatures of
the 4 sources span 300--600 K, which is lower than 1100--1300 K found
for the T Tauri sample of \citet{Najita03} but within the temperature
interval of 250--800 K given in \citet{Salyk09}. The discrepancy with
\citet{Najita03} may arise from differences in
methodology. \citet{Najita03} fit $^{12}$CO data covering lines with
upper energies levels of up to 10$^4$ K, which they assume to be
optically thin. They performed a linear fitting based on this
assumption, but curvature in their rotation diagrams indicates that
their $^{12}$CO lines are likely also optically thick, leading to
uncertainties in their extracted temperatures. Another explanation for
the temperature difference may be the lack of higher $J$-transitions
in our data set. However, AS 205 A includes $^{13}$CO transitions up
to $E_i$ = 4400 K, and still has a lower temperature of 550 $\pm$ 40
K. The temperature estimates by \citet{Salyk09} based on $^{12}$CO
should be more reliable since they took optical depth effects into
account. In addition, a rotational temperature of 770 $\pm$
  140 K has been estimated for DR Tau using the C$^{18}$O lines, which
  is close to the temperature of 510 $\pm$ 40 K given by the $^{13}$CO
  lines (Fig.~\ref{fig:13COrot}). If the two highest-$J$ $^{13}$CO lines are excluded from the fit, the $^{13}$CO rotational temperature increases to 610 $\pm$ 70 K, which is consistent with the C$^{18}$O temperature within the error bars. These two high-$J$ $^{13}$CO lines are very weak and their fluxes may have been underestimated.
 
We determine the $^{13}$CO/C$^{18}$O ratio for DR Tau to examine
whether the $^{13}$CO emission is indeed optically thin. The ratio is
calculated by comparing the fluxes in the R(5) and P(16) lines for
$^{13}$CO with the R(5) and P(17) lines for C$^{18}$O, chosen because
they are not blended and have similar $J$. This gives
  $^{13}$CO/C$^{18}$O = 6.3 $\pm$ 1.0 which is close to the overall
  abundance ratio of $^{13}$CO/C$^{18}$O = 8.5 observed in the solar
  neighborhood \citep{Wilson94}. The optical depth of $^{13}$CO is
  estimated to be $\tau$ = 0.3 $\pm$ 0.2. Inferred column densities
  and emitting areas will be presented in Brown et al.\ (in prep.),
  based on combined fits to the $^{12}$CO and $^{13}$CO lines.

\subsubsection{Vibrational temperatures} 

\begin{figure*}[htb]
\centering
{
 \includegraphics[width=140mm, angle=0]{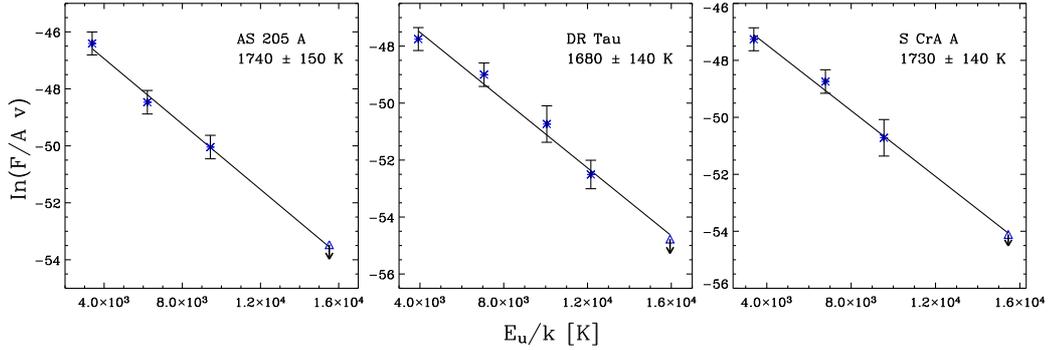}
 \caption{Vibrational diagrams for AS 205 A, DR Tau and S CrA A. Triangles represents upper limits.}
 \label{fig:drtau_vib}
}
\end{figure*}

Vibrational temperatures are determined for the three sources, AS 205
A, DR Tau and S CrA A, for which high $S/N$ $^{12}$CO data exist for
several vibrational transitions (see Fig.~\ref{fig:drtau_vib}). The
vibrational temperatures are calculated using the same relation
between the flux of the lines and the upper energy levels as presented
in equations \ref{eq:bol} and \ref{eq:Ni}. The assumption made here is
that every vibrational level has the same rotational
temperature. The only difference is that 
$\ell$n$(F_i$/$A_{ij}\nu$) instead of
$\ell$n$(F_i$/$A_{ij}\nu{g}_{i}$), is plotted vs the upper energy
levels $E_{i}$ to extract the vibrational temperature $T_{\rm vib}$ since the statistical weights of the vibrational levels are the same. One clear line, close in $J$-level and without strong line
overlap, is chosen per ro-vibrational level (CO $\rm{v} = 1-0$,
$\rm{v} = 2-1$ etc.) to extract the line fluxes.

The inferred vibrational temperatures are 1740 $\pm$ 150 K for AS 205
A, 1680 $\pm$ 140 K for DR Tau and 1730 $\pm$ 140 K for S CrA A. The
same fits have been done without including the $^{12}$CO $\rm{v} =
1-0$ lines since the lines within this band can be more optically
thick than those for the higher vibrational transitions. However,
excluding the $^{12}$CO $\rm{v} = 1-0$ lines does not significantly
change the results within the uncertainties. The upper limits of the
$\rm{v} = 5-4$ flux have been estimated using $F$ = 1.065 $I_{peak}$
$\Delta V$, where the $\Delta V$ is taken to be the same as that
of a CO $\rm{v} = 3-2$ line and the amplitude $I_{peak}$ =
3$\sigma$/$bin^{-1/2}$, where the parameter $bin$ is the number of
pixels within $\Delta V$ and $\sigma$ is the rms of a clean part
of the spectrum where the line of interest is expected. The detection
level is set to 3$\sigma$.

The vibrational temperatures of $\sim$1700 K are in general higher
than the rotational temperatures derived from the $^{13}$CO lines of
300-600 K. The inferred temperatures can be compared to
\citet{Brittain07} who find a lower rotational excitation temperature
of $\sim$200 K compared with a vibrational temperature of 5600 $\pm$
800 K for the disk around the Herbig Ae star HD 141569. Similarly,
van der Plas et al. (subm.) detect lower rotational temperatures
of $\sim$1000 K relative to vibrational values of $\sim$6000 -- 9000 K
for three other Herbig Ae/Be stars. Both papers interpret the higher
vibrational temperatures as caused by UV-pumping into the higher
vibrational levels.

The results presented here indicate that UV-pumping may also
  be an important process for populating the higher vibrational levels
  in disks around T Tauri stars, especially for the broad-based single
  peaked sources. The rotation/vibration temperature differences in
  \citet{Brittain07} and van der Plas et al. (subm.) are, however, much larger than those derived
  here. This may be due to the stronger UV fields from Herbig Ae/Be
  stars than T Tauri stars in the wavelength range where CO is
  UV-pumped. 

  Additional UV flux above the stellar photosphere can be produced by
  accretion \citep{Valenti00}. Table \ref{tab:acc_peak} includes
  accretion luminosities taken from the literature for our T Tauri
  stars. These accretion luminosities are based on measurements of the
  hydrogen continua or H$\alpha$ line emission from the sources. The
  average accretion luminosity for those broad-based single peaked
  lines for which accretion luminosities have been measured, is 0.5
  L$_\odot$. This value is higher than the average accretion
  luminosity of 0.1 L$_\odot$ which is calculated for the rest of the
  sample of disks that are shown in Table \ref{tab:acc_peak}. The
  strong Pfund-$\beta$ lines observed in our data provide an
  additional confirmation of ongoing strong accretion (see
  Fig.~\ref{fig:source_setting}). Higher accretion rates lead to
  higher UV fluxes, which, in turn, increase the UV pumping rate of CO
  into higher vibrational states.  Quantitative estimates of the
  effect have been made by Brown et al.\ (in prep.) using the actual
  observed UV spectra for stars with spectral types A--K in
  an UV excitation model. Even if the enhanced UV from accretion for
  the T Tauri stars is included, however, the resulting vibrational
  excitation temperatures for a K-type star with additional UV
    due to accretion are still lower than those using an A-star
  spectrum. For the ${\rm v}=1-0$ emission, the question remains how
  much UV-fluorescence contributes relative to thermal excitation.

\subsection{Lack of extended emission}\label{extended_sec}

\begin{table*}
\footnotesize
\caption{The upper limit on the projected 
radial extent $\Delta{R_{line}}$ of the line emission, together with 
$\Delta{R_{instr}}$.}
 \begin{minipage}[t]{\columnwidth}
  \renewcommand{\footnoterule}{} 
 \centering
  \thispagestyle{empty}
  \begin{tabular}{l c c}
   \hline
    \hline 
   Source & $\Delta{R_{instr}}$\footnote{See text for definition} [AU] & $\Delta{R_{line}}$ [AU] \\
   \hline
AS 205 A & 15.1 $\pm$ 0.1 & $<$2.1 \\
DR Tau & 14.5 $\pm$ 0.1 & $<$1.7 \\
RU Lup & 17.9 $\pm$ 0.2 & $<$2.5 \\
S CrA A & 15.9 $\pm$ 0.2 & $<$2.0 \\
S CrA B\footnote{Contamination by companion star.} & - & - \\
VV CrA A$^{b}$ & 16.6 $\pm$ 0.5 & - \\
VW Cha & 23.4 $\pm$ 2.3 & $<$10.4\\
VZ Cha & 20.7 $\pm$ 1.0 & $<$6.3 \\
  \hline
\end{tabular}
\end{minipage}
\label{tab:sum_2ndmom}
\end{table*}

\begin{figure}[htb]
\centering
{
 \includegraphics[width=80mm]{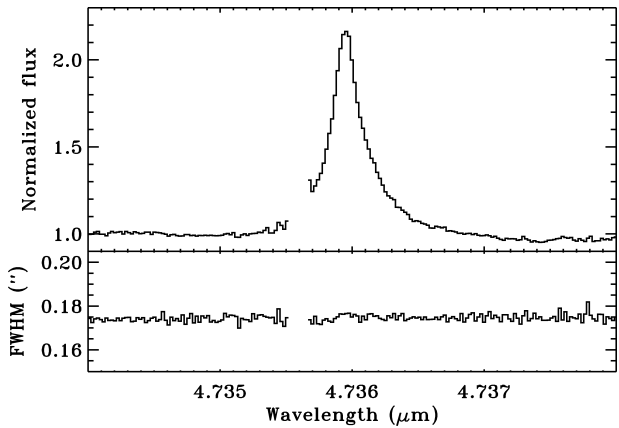}
 \caption{Top: The spectral line CO $\rm{v} = 1-0$ P(8) for AS 205. Bottom: The FWHM of the spatial profile of the above spectrum at each individual wavelength point. No increase in FWHM is seen at the position of the line.}
 \label{fig:FWHM}
}
\end{figure}

\begin{figure*}[htb]
\centering
{
 \includegraphics[width=70mm, angle=0.0]{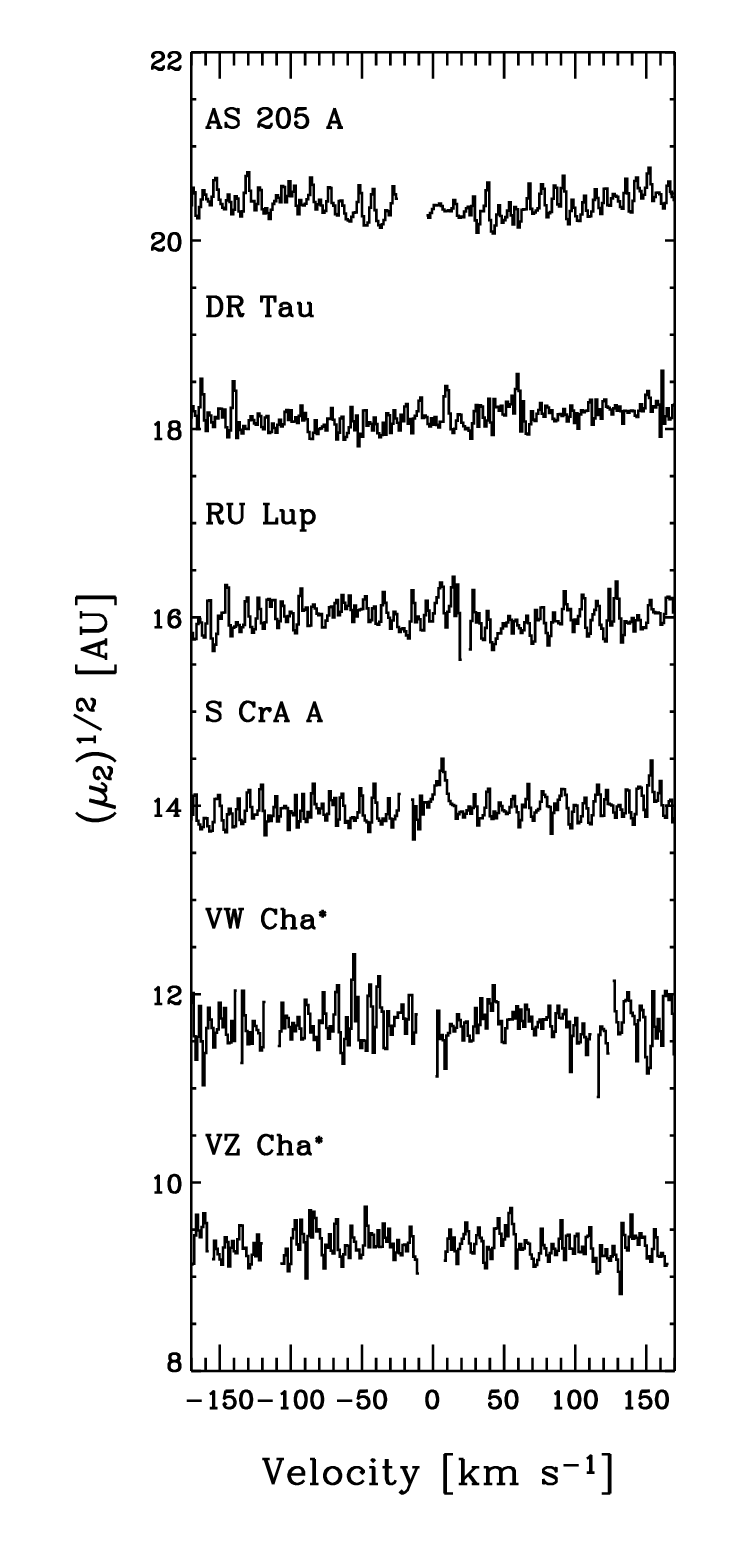}
 \caption{The stacked square root of the formal second moment $(\mu_2)^{1/2}$ of the spectral
   trace for the $^{12}$CO emission lines. About 5--10 clean lines for each source were used for stacking. The square root of the second moment $(\mu_2)^{1/2}$
   is multiplied with * = 0.2 for sources
   with a larger uncertainty in the second moment. The vertical shift for respective source is: AS 205 A = 0 AU, DR Tau = -3.4 AU, RU Lup = -7.0 AU, S CrA A = -7.2 AU. For VW Cha and VZ Cha the shifts are 5.5 and 3.7 AU after multiplication with 0.2.}
 \label{fig:2ndmom}
}
\end{figure*} 

Information about the spatial extent of the CO emission from
broad-based single peaked sources is an important ingredient for
constraining models of their origin. The variance or formal second
moment $\mu_2$ of the spectral trace is calculated at each wavelength
from the 2-dimensional spectrum, according to the following equation.

\begin{equation}
\mu_2 = \Sigma (x_i - C)^2 \times F_i / \Sigma F_i
\end{equation}
where $x_i$ is the spatial position, $F_i$ is the flux at
that point and $C$ is the centroid position computed from
$C = \Sigma x_i \times F_i / \Sigma F_i$. 

For continuum emission, the average value of the second moment
reflects the spatial resolution provided by the AO system, assuming
that the continuum is not spatially extended. If the line emission is
spatially extented, the values of $\mu_2$ will be larger than the
noise in the continuum $\mu_2$ signal at wavelengths where the line
emits. There is no significant detection of extended line emission for
any of our sources, see Fig.~\ref{fig:FWHM} and
Fig.~\ref{fig:2ndmom}. An upper limit on the radial
  extent of the line emission, $\Delta{R_{line}}$, is therefore
  calculated using the formula $\Delta{R_{line}}^2$ $\leq$
  ($\Delta{R_{instr}} + \sigma_{\Delta R}$)$^2$ $-
  \Delta{R_{instr}}^2$, where $\Delta{R_{instr}^2}$ is equal to the
  mean of $\mu_2$/2 of the continuum and $\sigma_{\Delta R}$ is the standard deviation of
  ($\mu_2$/2)$^{1/2}$. The values listed in Table \ref{tab:sum_2ndmom}
  include an additional correction factor (1 + $C/L$), where
   $C/L$ is the continuum/line flux ratio, because the total centroid is
  diluted by the continuum flux. These results show that most of the
projected CO emission must originate from within a few AU for
AS 205 A, DR Tau, RU Lup and S CrA A. S CrA B and VV CrA A (not shown)
have a feature close to the central velocity of the line, however this
can be explained by contamination from the companion star. The small
feature seen for S CrA A may be a detection of extended line
emission. VW Cha and VZ Cha have higher upper limits of the
projected radial line extent of 10.4 and 6.3 AU respectively
which is caused by a higher noise in the second moment of the
continuum emission.

Since the position angle of the slit is arbitrary with respect
  to that of the disks in our observations, an additional check on
the lack of radial extended emission is provided by the
different spatial line profiles of the stacked CO $\rm{v} = 1-0$ lines
of AS 205 A, DR Tau and S CrA A taken at 3 different rotational
angles.  All three show exactly the same profile. This means that the
CO emission originates from within a few AU at many different angles
around the sources.

\section{Discussion}  \label{discussion}

Three possibilities for the origin of broad-based single peaked
profiles are discussed below: a Keplerian rotating disk, a disk wind
or a funnel flow.

\subsection{Rotating disk} \label{disc_rot_disk}

\begin{figure*}[htb]
\centering
{
 \includegraphics[width=90mm, angle=0.0]{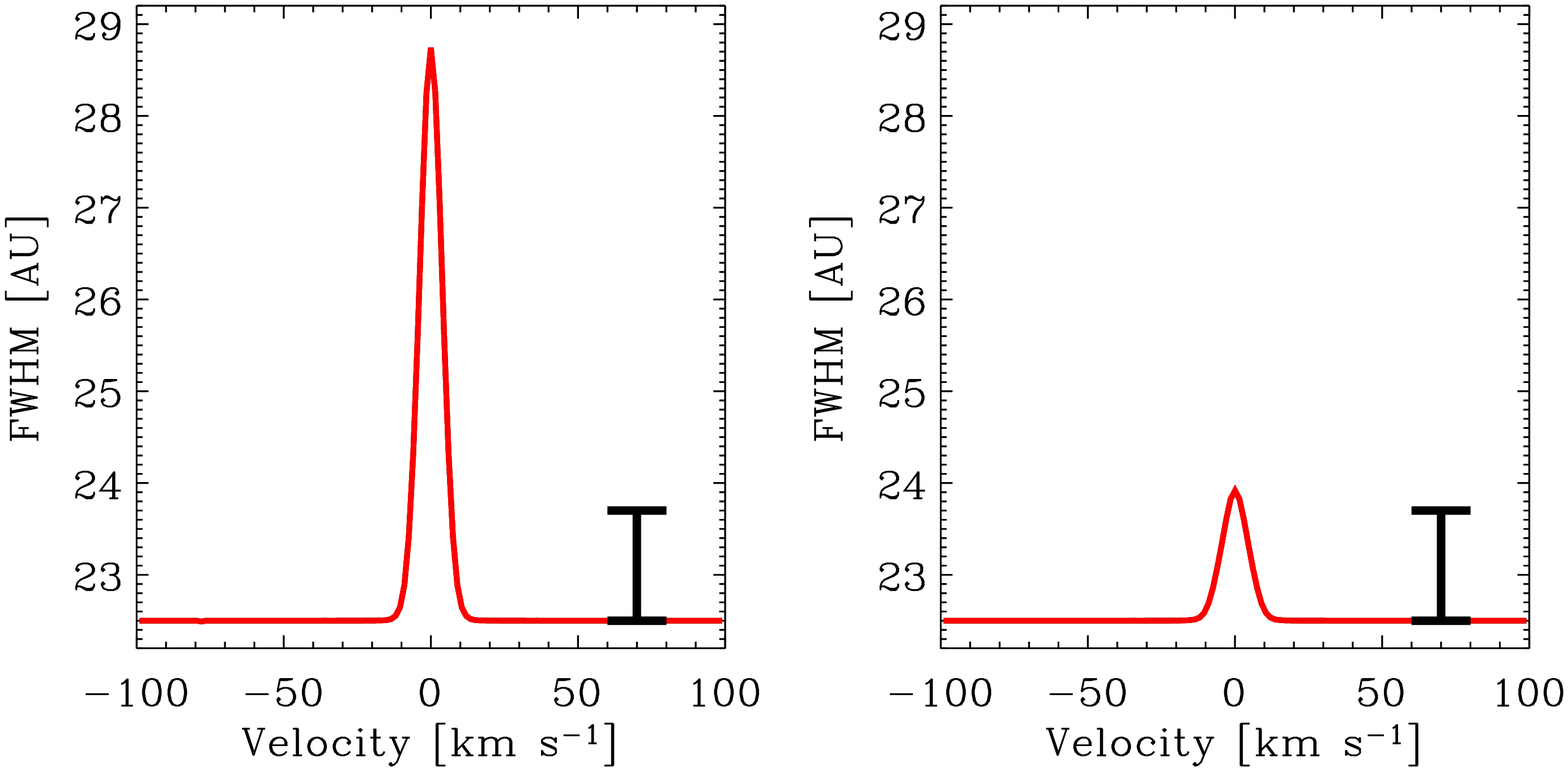}
 \caption{The second moment of the modeled spatial line profile for AS
     205 (10 AU = 0$\farcs$08 at a distance of 125 pc). The model in the left plot includes extended emission out to
     30 AU and $\Delta V_{\rm mod}$=3 km s$^{-1}$. The model in the
     right plot includes emission out to 10 AU and an increased
     $\Delta V_{\rm mod}$=7 km s$^{-1}$ (see
     Fig.~\ref{fig:iterative}). Error bars represent the 3$\sigma$ noise limit for AS 205 A in Fig.~\ref{fig:2ndmom}. Both models
     use the observed position angle of 115$\degr$ which is 50$\degr$
     off from the position angle of 165$\degr$ of the AS 205 A disk
     \citep{Andrews09}.} \label{fig:iterative2} }
\end{figure*}

The results presented in $\S$\ref{keplerian} show that
  broad-based single peaked lines (category C) cannot be reproduced
  using a Keplerian model with a power-law temperature profile. Increasing the local line
  broadening parameter $\Delta{V}_{\rm mod}$ to $\sim$8 km s$^{-1}$
  removes the double peak, but a larger local line broadening is not
  sufficient to give a good fit between the disk model and the
  data. As seen in $\S$ \ref{iterative} a Keplerian disk model can fit the data by solving for a temperature structure. The best fits have a plausible temperature structure but require extended emission, possibly in combination with an increase in the local broadening parameter.
  
   The results of \S~\ref{extended_sec} show that the bulk of
     the emission comes from within a few AU which means that a
     Keplerian model with a non-standard temperature profile including extended emission out to 30 AU cannot explain the broad-based single peaked line profiles. Even
     the model with enhanced turbulence and emission out to 10 AU
     becomes marginal, as illustrated by the expected second moments
     in these non-standard cases (Fig.~\ref{fig:iterative2}). The
     vertical error bars in Fig.~\ref{fig:iterative2} represent the
     3$\sigma$ noise limit in AU for AS 205 A.
     Similar results hold for most of the other sources, although the
     Keplerian models with a non-standard temperature profile including an enhanced turbulence cannot be fully
     ruled out for sources like VW Cha and VZ Cha for which
     the limits on the spatial extent are larger.

   \citet{Carr04} invoked a larger local line broadening of
     7--15 km s$^{-1}$ in their disk models to obtain a better fit to
     CO $\rm{v} = 2-0$ band head profiles observed toward various
     young stellar objects. As discussed in their paper, such values
     are much larger than the sound speed of a few km s$^{-1}$ in the
     inner disk, whereas models of turbulence in disks typically give
     velocities that are less than the sound speed
     \citep[e.g.,][]{Klahr06}. In addition, these turbulent values are
     much higher than those inferred from observations of gas in the
     outer disk \citep{Qi06,Hughes09}, although the turbulence in a
     warm surface layer near the star may differ from that in the
     outer regions.  A more extended discussion of turbulence in inner
     disks is given in \citet{Carr04}. Even though we cannot fully
     exclude an increased level of turbulence for some of our sources,
     this is unlikely to be the sole explanation for the broad-based
     single peaked line profiles observed here.

 In summary, the above discussion shows that a pure Keplerian disk model
   does not fit the data. However, the estimated rotational temperatures of
 300 to 800 K for the broad centrally peaked sources are consistent
 with the temperature expected for the inner parts of a disk. The
 symmetry of the lines and the small radial velocity shift between the
 gas and the star (a maximum of $\sim$5 km s$^{-1}$) are additional
 indications that the emission may originate in a
 disk. Therefore, it is likely that at least part of the
   emission originates from the disk but some additional process
 must contribute to these line profiles as well.

\subsection{Disk wind}
\subsubsection{Thermally launched winds}

Strong outflows are excluded due to the small velocity
shifts of the molecular gas relative to the central star. Thermally
launched winds are driven by irradiation of either FUV
(far-ultraviolet), EUV (extreme ultraviolet), or X-ray photons
\citep{Alexander08,Gorti09_2,Ercolano10}. Irradiation of the upper
layers of the disk heats the gas to such high temperatures that the
thermal energy exceeds the gravitational potential, leading to a flow
off the disk surface. The launching radius is roughly given by $R_g$ =
$GM_\star/V_{\rm{orb}}^2$, where $V_{\rm{orb}}$ is the orbital speed
of the gas. The wind retains the Keplerian rotational velocity
of the launch radius.

If the disk is illuminated by EUV radiation, an HII region forms in
the upper layer of the disk atmosphere. The ionized gas reaches a
temperature of 10$^4$ K, which corresponds to $V$ $\sim$10 km
s$^{-1}$. For a 1 $M_\odot$ star, the launching radius is then
$R_g$$\sim 5 - 10$ AU, and the wind is launched with a velocity of $\sim
5-10$ km s$^{-1}$ \citep{Alexander08}. This small velocity shift is
consistent with the small velocity shifts we see in the CO gas
emission relative to the stellar radial velocity. However, an
EUV-driven wind consists mostly of ionized gas, so that CO may be
photodissociated. Detailed models of the CO chemistry are needed to
test whether enough CO can survive in such a wind to produce the
observed CO line profiles.

In the X-ray and FUV photoevaporation models of \citet{Gorti09_2}, the
evaporating gas is cooler than the EUV-irradiated gas. The expected
launching radii of 10--50 AU for X-ray photoevaporation and $>$50 AU
for FUV photoevaporation are inconsistent with the lack of
spatially-extended emission in our CO spectra. However, in models of
photoevaporation by soft X-ray emission by \citet{Ercolano10}, the
launching radius is closer to the star. \citet{Ercolano10}
calculate profiles for atomic and ionized species within this model
and predict symmetric, broad-based, single-peaked line profiles with
small blueshifts of 0--5 km s$^{-1}$.  

As for the EUV models, a detailed CO chemistry needs to be coupled
with the wind model to test this scenario, but the fact that the wind
is mostly neutral will help in maintaining some CO in the flow.

In summary, EUV and soft-X-ray photoevaporation of the disk might produce some CO emission at low velocities.  More detailed studies of the CO chemistry and velocity fields in the thermally-launched wind are needed to determine where CO can survive in the wind and the
resulting emission line profile.

\subsubsection{Magnetically launched winds}

Magnetic fields of Classical T Tauri stars exert torques that are
thought to be strong enough to launch powerful winds.
\citet{Edwards06} describe two types of winds that are seen in
absorption in the \ion{He}{1} $\lambda 10830$ line, one with
velocities of around 250 km s$^{-1}$ and another with lower speeds up
to 50 km s$^{-1}$. These velocity shifts are also commonly observed in
optical forbidden lines, such as [O I] 6300 \r{A}. The high velocities
are not consistent with the velocities seen in the broad-based single
peaked CO emission, making this scenario highly unlikely.

\subsection{Funnel flow}

Accretion from the disk onto the star is thought to occur in a funnel
flow along strong dipolar field lines
\citep[e.g.,][]{Hartmann1994,Bouvier07,Yang07,Gregory08}.
\citet{Najita03} rule out the possibility that the funnel flow could
produce some CO emission because the temperatures in the funnel flow
are expected to be $\sim 3000-6000$ K \citep{Martin1997}, higher than
the measured temperatures of the CO-emitting gas. Models of observed
hydrogen emission line profiles are consistent with these temperatures
\citep{Muzerolle01,Kurosawa06}. However, \citet{Bary08} suggest that
the relative hydrogen emission line fluxes may also be consistent with
cooler temperatures of $\sim 1000-1500$ K, and cooler temperatures may
be present where the funnel flow connects with the
disk. \citet{Najita03} show that a modeled CO emission line profile
from a funnel flow would have a double peaked line profile with no
velocity shift (see Fig. 10b in their paper), which disagrees with the
presence of the narrow emission peak seen in the line profiles
discussed here. Thus, funnel flows are also not likely to be
  the origin of the observed emission.

\section{Conclusions}\label{conclusions}

Using CRIRES spectroscopy of CO ($\Delta \rm{v} = 1$) ro-vibrational
emission lines around 4.7\,$\mu$m from a sample of 50 T Tauri stars,
we find that the line profiles can be divided into three basic
categories: A) narrow single peaked profiles, B), double peaked
profiles and C) single peaked profiles with a narrow peak ($\sim10 -
20\,\rm km\,s^{-1}$), but broad wings, sometimes extending out to
100\,$\rm km\,s^{-1}$. The broad-based single peaked line profiles are
rather common, since they were detected in 8 sources in a sample of
about 50 T Tauri stars. These 8 sources have preferentially high
accretion rates, show detections of higher vibrational
  emission lines (up to $v = 4$) and have CO lines with
unusually-high line-to-continuum ratios relative to all other sources
within our sample. For at least one of the disks (S CrA B),
  the narrow component decreases faster for higher $J$ lines than the
  broad component, implying that the narrow component is colder.
  Generally, however, the two component fits give similar temperatures
  and central velocities.

The broad-based single peaked emission originates within a few AU of
the star with gas temperatures of $\sim$300 -- 800 K. The line
profiles are symmetric and the gas radial velocity is close to the
stellar radial velocity. These characteristics point toward an origin
in a disk. However, unlike the CO profiles from other objects that are
double-peaked or have only narrow peak, the broad-based single peaked
line profiles could not be well fit using a Keplerian model with a
power-law temperature profile. The fits are improved if a large
turbulent width of $\sim8$ km s$^{-1}$ is invoked, but the overall
profile fit is still poor.  Models with an unusual temperature
distribution, perhaps with enhanced turbulence, provide a much better
fit to the line profiles but require emission out to larger distances
than observed.  Thus, the hypothesis that this emission originates in
a pure Keplerian disk needs to be questioned and an additional
physical component needs to be considered.

Several other scenarios are also ruled out. FUV radiation-driven
winds have a launching radius of $>50$ AU that is inconsistent with
the lack of spatially-extended emission in our data.
Magneto-centrifrugal winds are observed to have blueshifts of 50 -- 250
km s$^{-1}$, which are inconsistent with the lack of a significant
velocity shift in the symmetric CO emission lines. A funnel flow is
also unlikely because emission near the inner rim of the disk should
have a double-peaked line profile, and because the temperatures in the
funnel flow are expected to be higher than the calculated CO
rotational temperatures. 

The most plausible explanation for the broad-based single-peaked line
profiles is therefore some combination of emission from the warm
surface layers of the inner disk, contributing to the broad
  component through Keplerian rotation, and a disk wind, responsible
  for the narrow component. A thermally launched disk wind, perhaps
driven by EUV radiation or soft X-rays will have a small velocity
shift of $\Delta{V}$ of 0-10 km s$^{-1}$, which is marginally
consistent with the maximum detected velocity shift of
$\Delta{V}\sim$5 km s$^{-1}$.

More information, including independent estimates of inclination
angles, detailed models of CO chemistry and line profiles
that originates in disk winds and the inner regions of disks, and
spectro-astrometry of emission from these sources, is needed to
distnguish the contributions from a disk (perhaps with enhanced turbulence) 
and a disk wind. These line profiles will be further discussed by Pontoppidan,
Blake \& Smette (subm.) using a combined disk and slow molecular
disk wind model to analyze the spectro-astrometric data of these
sources.

  \begin{acknowledgements}
  
    JEB is supported by grant 614.000.605 from Netherlands
    Organization of Scientific Research (NWO). EvD acknowledges
    support from a NWO Spinoza Grant and from Netherlands Research
    School for Astronomy (NOVA). JEB is grateful for the hospitality
    during long term visits at the Max Planck Institute for
    Extraterrestrial Physics in Garching and Division of Geology and
    Planetary Science at California Institute of Technology in
    Pasadena. The authors would like to acknowledge valuable
    discussions with R. Alexander, B. Ercolano, C. Salyk,
    J. Muzerolle, A. Johansen, A. Smette, U. K\"aufl and W. Dent.

\end{acknowledgements}

\end{document}